\documentclass[twocolumn,10pt,nofootinbib]{revtex4}

\def\mysections#1{{\bf #1.} }

\usepackage[dvips]{graphicx}
\usepackage{epsfig,amsmath,amssymb,verbatim,mathrsfs,array,layout,textcomp,amssymb,xcolor,latexsym,slashed,graphicx,bbm}

\usepackage{graphicx}
\usepackage{amsmath}
\usepackage{amssymb}
\usepackage{appendix}
\usepackage{verbatim,mathrsfs,array,layout,textcomp,latexsym, slashed}
\usepackage{float}

\usepackage{amsmath, amssymb, graphicx, rotating}
\usepackage{hyperref}

\newcommand{\mean}[1]{\left\langle #1 \right\rangle}
\newcommand{\R}{\mathcal{R}}
\newcommand{\PP}{\mathcal{P}}
\newcommand{\bt}{\boldsymbol{t}}
\newcommand{\xvec}{\vec{x}}

\newcommand{\beq}{\begin{eqnarray}}% can be used as {equation} or {eqnarray}
\newcommand{\eeq}{\end{eqnarray}}

\def\beqa{\begin{eqnarray}}
\def\eeqa{\end{eqnarray}}

\newcommand{\bv}{\left(\begin{array}{c}}
\newcommand{\ev}{\end{array}\right)}
\newcommand{\emn}{\end{array}\right)}
\newcommand{\bmtwoc}{\left\{\begin{array}{cc}}
\newcommand{\bmthreec}{\left\{\begin{array}{ccc}}
\newcommand{\emnc}{\end{array}\right\}}
\newcommand{\ba}{\begin{array}}
\newcommand{\ea}{\end{array}}

\def\lsim{\mathrel{\rlap{\lower4pt\hbox{\hskip1pt$\sim$}}
     \raise1pt\hbox{$<$}}}         %less than or approx. symbol
\def\gsim{\mathrel{\rlap{\lower4pt\hbox{\hskip1pt$\sim$}}
     \raise1pt\hbox{$>$}}}         %greater than or approx. symbol

\begin{document}

\font\mini=cmr10 at 0.8pt

\title{
Gravitational Waves from  Incomplete Inflationary Phase Transitions}

\author{Joel Barir}
\email{joelbari@mail.tau.ac.il}
\author{Michael Geller}\email{micgeller@tauex.tau.ac.il}
\author{Chen Sun}\email{chensun@mail.tau.ac.il}
\author{Tomer Volansky}\email{tomerv@post.tau.ac.il}
\affiliation{School of Physics and Astronomy, Tel Aviv University, Tel Aviv, Israel}

%\date{\today}
\begin{abstract}
We study the observable implications of an incomplete first order phase transition during inflation. 
In such a phase transition, the nucleated bubbles  do not percolate and instead are continuously produced until the onset of reheating. The process creates an inhomogeneity with a distinct power spectrum that depends on both the physics of the phase transition and 
the inflationary dynamics. 
Upon horizon re-entry, this spectrum generates gravitational waves through non-linear effects. This stochastic gravitational wave background is predicted to have unique signatures that may be detectable by future experiments spanning a wide frequency range.  The discovery of such a gravitational wave signal would shed a light on the detailed dynamics of inflation.
\end{abstract}

\maketitle

%%%%%%%%%%%%%%%
\section{Introduction}
The  detection of gravitational waves (GWs) by LIGO and VIRGO
\cite{LIGOScientific:2016aoc} has opened a new window into our universe.
Upcoming and far future experiments are expected to cover a wide range of frequencies and also improve current sensitivities
\cite{KAGRA:2013rdx,amaroseoane2017laser,TianQin:2015yph,Crowder:2005nr,Harry:2006fi,Corbin:2005ny,Kramer:2013kea,2010, Janssen:2014dka,Punturo:2010zz,Reitze:2019iox}, making the future detection of
a stochastic gravitational wave background possible.  Intriguingly, the NANOGRAV collaboration has
reported some hints for the existence of a stochastic gravitational
wave background at low frequencies of order 1 yr$^{-1}$
\cite{NANOGrav:2020bcs}, although it is not clear if the origin is cosmological.
Since gravitational waves propagated freely through the universe even while it was opaque to light, a gravitational wave background of
primordial origin, if it exists, can hold crucial information about the very early history of our universe.

Inflation is the leading paradigm for these first moments~\cite{Guth:1980zm,Linde:1981mu}.
A period of rapid, exponential expansion explains why the observable universe
is flat, homogeneous and isotropic to a very good accuracy.  In the simplest
scenario, the expansion can be driven by a single scalar field, called the
inflaton. The initially small inhomogeneities of the universe originated in inflationary
vacuum fluctuations, which grew to cosmological scales.  These inhomogeneities are observed through the cosmic microwave background (CMB), giving strong constraints on
inflationary models~\cite{Planck:2018jri,BICEP2:2018kqh,WMAP:2008lyn,Carr:2020gox,1996ApJ...473..576F}. For a review of inflation, see
\cite{baumann2012tasi}.

GWs provide a promising tool to further explore the inflationary epoch. Although nothing is known about the pre-inflationary epoch, it is commonly assumed that the universe has been in a high-energy state. During inflation, the universe has rapidly cooled and it is natural to expect the system to be away from its global minimum, which may be eventually reached through one or more phase transitions (PTs). It is conceivable that some of these PTs are first order and proceed through bubble nucleation.
Fast enough first order PTs proceed through a percolation stage which produce (possibly observable) GWs~\cite{An:2020fff,Wang:2018caj,An:2022cce} (for GWs from PTs after inflation, see, e.g.~\cite{Kosowsky:1992vn,Kosowsky:1991ua,Kamionkowski:1993fg,Huber:2008hg,Caprini:2007xq,Caprini:2015zlo,Caprini:2019egz,Hindmarsh:2013xza,Giblin:2014qia,Hindmarsh:2015qta,Kahniashvili:2008pf,Kahniashvili:2009mf,Caprini:2009yp,Schmitz:2020syl}).  
Another possibility, however, which is at the focus of this letter, is slow PTs, which do not complete during inflation.  In that case, bubbles nucleate too far from each other and never go through the percolation stage, due to the shrinking Hubble sphere. One may therefore wonder whether a GW signal still forms and if so, how would it be distinguished from the previously studied scenarios?  

As we show below, while bubbles do not collide and percolate, their presence serves as a new source of 
inhomogeneities on small length scales. As a consequence of the PT remaining incomplete during inflation, the bubbles are produced continuously, resulting in a broad and flat inhomogeneity spectrum, spanning across a large range of modes. 
After inflation, once those modes enter the horizon, the inhomogeneities induce GWs through secondary effects~\cite{Matarrese:1997ay,Mollerach:2003nq,Ananda:2006af,Baumann:2007zm,Kohri_2018} with a similarly broad and rather unique spectrum which could be measured by multiple upcoming and future experiments. The spectrum depends not only on the sector which goes through the PT but also on the details of inflation.  

To demonstrate the above, we study a simplified single-field model which captures the relevant features of various PTs, including the well-known Coleman-de Luccia (CdL) bubble nucleation~\cite{ColemanDeLuccia} and the Hawking-Moss (HM) instanton~\cite{Hawking:1981fz}.  We then calculate the expected anisotropies and resulting GW background, concluding that a new and promising signal may appear in future GW observatories, shedding light on hidden sectors as well as on the physics of inflation.

%%%%%%%%%%%%%%%%
\section{Inflation and Phase Transitions}\label{sec:HM}
A first order PT in the early universe takes place via bubble nucleation.   The bubbles may or may not collide depending on the competition between their expansion and nucleation rates with the expansion rate of the universe.  In this sense, such phase transitions exhibit two distinct regimes. The phenomenology of PTs with bubble collisions has been thoroughly studied~\cite{An:2020fff,Wang:2018caj,An:2022cce,Kosowsky:1992vn,Kosowsky:1991ua,Kamionkowski:1993fg,Huber:2008hg,Caprini:2007xq,Caprini:2015zlo,Caprini:2019egz,Hindmarsh:2013xza,Giblin:2014qia,Hindmarsh:2015qta,Kahniashvili:2008pf,Kahniashvili:2009mf,Caprini:2009yp,Schmitz:2020syl}, and here we focus on the signatures of PTs where the bubbles can't meet.

During inflation, any slow enough first order PT does not complete~\cite{GUTH1983321,Turner:1992tz}. Schematically, if the bubble nucleation rate per unit volume is smaller than the Hubble expansion rate, i.e.,
\begin{equation} \label{eq:slow}
\Gamma/V\lesssim H^{4},
\end{equation} 
the mean distance between two neighboring bubbles is larger than the cosmological horizon (which shrinks in comoving coordinates). Hence bubbles cannot meet and percolate, leaving most of space in the false vacuum and the transition incomplete for as long as the universe inflates. While signals, such as GWs, are typically known to be produced during the percolation period, in this paper we show that stochastic GW signals are also predicted in slow PTs that can't percolate during inflation, and the resulting signal records the entire duration of the PT. 

To be concrete, consider first the CdL tunneling process~\cite{ColemanDeLuccia} which describes the 
quantum process of vacuum tunneling in a gravitational background, and the rate of which is calculated with the instanton method based on the saddle-point approximation. 
In the semiclassical calculation, the bubbles are produced at rest with their radius equal to the critical radius - the minimal radius for an expanding bubble. 
Once formed, such bubbles expand classically, quickly approaching the
speed of light.
As soon as the physical
radius of the bubble becomes larger than the Hubble radius, $H^{-1}$,
the surface velocity becomes negligible and the Hubble drift
dominates the bubble evolution. At this point the bubble is 
``frozen", i.e.\ it does not expand with respect to the comoving frame.  As a consequence, a single bubble can never overtake the entire universe and for low enough nucleation rate, too few bubbles can form to complete the PT. 

%\subsection{The Hawking  Moss Transition}
The Hubble radius represents the region contained inside a 
cosmological horizon created by the expanding universe. 
Therefore, in order to maintain causality, a CdL bubble must form with a smaller radius and  if the critical bubble radius is larger than $H^{-1}$,  the CdL instanton does not exist.
Instead,  tunneling is still possible through the HM solution
\cite{Hawking:1981fz}. In the HM case, an entire Hubble 
patch tunnels simultaneously to the top of the potential barrier.  This phenomenon is best understood through the formalism of
stochastic inflation \cite{Starobinsky:1986fx} where the inflationary 
horizon gives rise to a temperature, analogous to the Hawking temperature
of a black hole. The thermal fluctuations then allow a trapped scalar field to 
diffuse, eventually reaching the top of a potential barrier. Once
the barrier is crossed, the field may classically roll to the true
minimum. 
As with the CdL PT, here too the Hubble-size bubble remains frozen and a slow nucleation rate implies that the PT is never complete.   We stress, however, that as opposed to the CdL case, the stochastic formalism shows that the HM instanton calculation  only holds in the limit of a very slow transition~\cite{Linde:2005ht}, and thus by construction can only describe an incomplete PT during inflation.

%%%%%%%%%%%%%%%%
\section{A Model}\label{sec:model}
The necessary details needed to study an incomplete PT can be described by a simple toy model.  The PT is driven by the field $\chi$, acting as a spectator during inflation. The inflationary dynamics are dominated by the inflaton $\phi$, for which we assume the slow-roll conditions to hold, but whose detailed potential we otherwise remain agnostic to.   While interactions between $\chi$ and $\phi$ may exist, their presence do not significantly affect our conclusions and we ignore them here. The potential is thus
\begin{equation}
	V = V_{\rm PT}(\chi) + V_{\rm inf}(\phi)\,,
\end{equation}
where,
\begin{equation}
V_{\rm PT}(\chi)\ll V_{\rm inf}(\phi)\,.
\end{equation} 

\begin{figure}[t]
	\centering
	\includegraphics[width=\columnwidth]{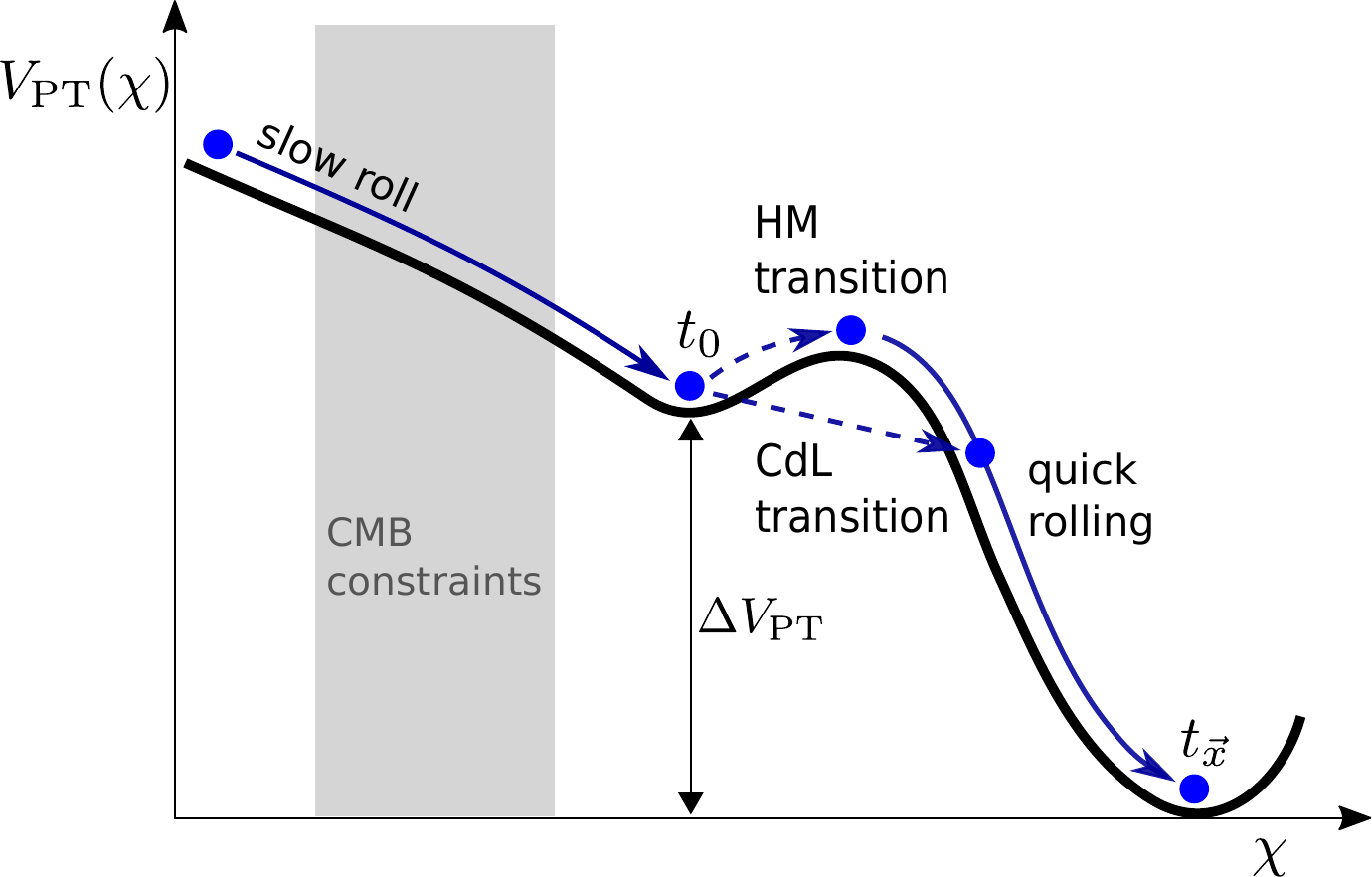}
	\caption{Evolution of the spectator field, $\chi$, that drives the phase transition. A slow-rolling phase allows for the creation of the almost scale-invariant power spectrum observed at large scales in the CMB. The region probed by the CMB is illustrated with the gray-colored region.  At some later time $t_0$, the field gets stuck in a local minimum of the potential. The field may escape this local minimum before the end of inflation through either Coleman-de Luccia (CdL) or Hawking-Moss (HM) tunneling, creating inhomogeneities in the energy density.  Under the assumption of low bubble nucleation rate, this phase transition is never completed during inflation.  We assume $V(\chi)\ll V(\phi)$ where $\phi$ is the inflaton field, for the entire duration of inflation.}
	\label{fig:model}
\end{figure}

The PT potential,  $V_{\rm PT}(\chi)$, is illustrated in Fig. \ref{fig:model}. We take $\chi$ to be initially on the left side of the potential and away from the false vacuum,  classically slow-rolling down. To evade CMB constraints~\cite{Planck:2018jri,BICEP2:2018kqh,WMAP:2008lyn,Carr:2020gox,1996ApJ...473..576F} we assume that $\chi$ settles at the unstable minimum sufficiently late in the inflationary epoch, so that bubble nucleation takes place only after the CMB modes had exited the horizon.
Once at the local minimum, $\chi$ can
tunnel over the potential barrier and roll to the global minimum where
$V_{\rm PT}=0$, nucleating true vacuum bubbles. For simplicity, we assume this classical rolling after barrier crossing to be instantaneous.

The bubble nucleation rate during the PT is directly dictated by the potential parameters.   We thus choose $V_{\rm PT}$  such that Eq.~\ref{eq:slow} is fulfilled, ensuring a  slow PT and implying that the physical volume of space where $\chi$ is ``stuck" in the unstable minimum increases with time. 
Once the inflaton decays, regions of false vacuum may dominate the energy density and lead to an unwanted eternal inflation within our Hubble patch, driven by $\chi$.   To evade such a catastrophe, one may either assume that the reheating temperature is larger than the energy density in the false vacuum and its effect drives to destabilize it, or even simpler, that the nucleation rate is larger than the value of Hubble in the false vacuum so that rapid nucleation and percolation becomes possible after the inflaton decays\footnote{We will ignore the GW from this final stage of the PT, as these occur not far from the reheating time, and the frequency range is likely beyond any near-future experiment.}.  With this, the PT suddenly and instantaneously completes everywhere and inflation truly ends at least within our visible universe.

\begin{figure}[t]
	\centering
	\includegraphics[width=\columnwidth]{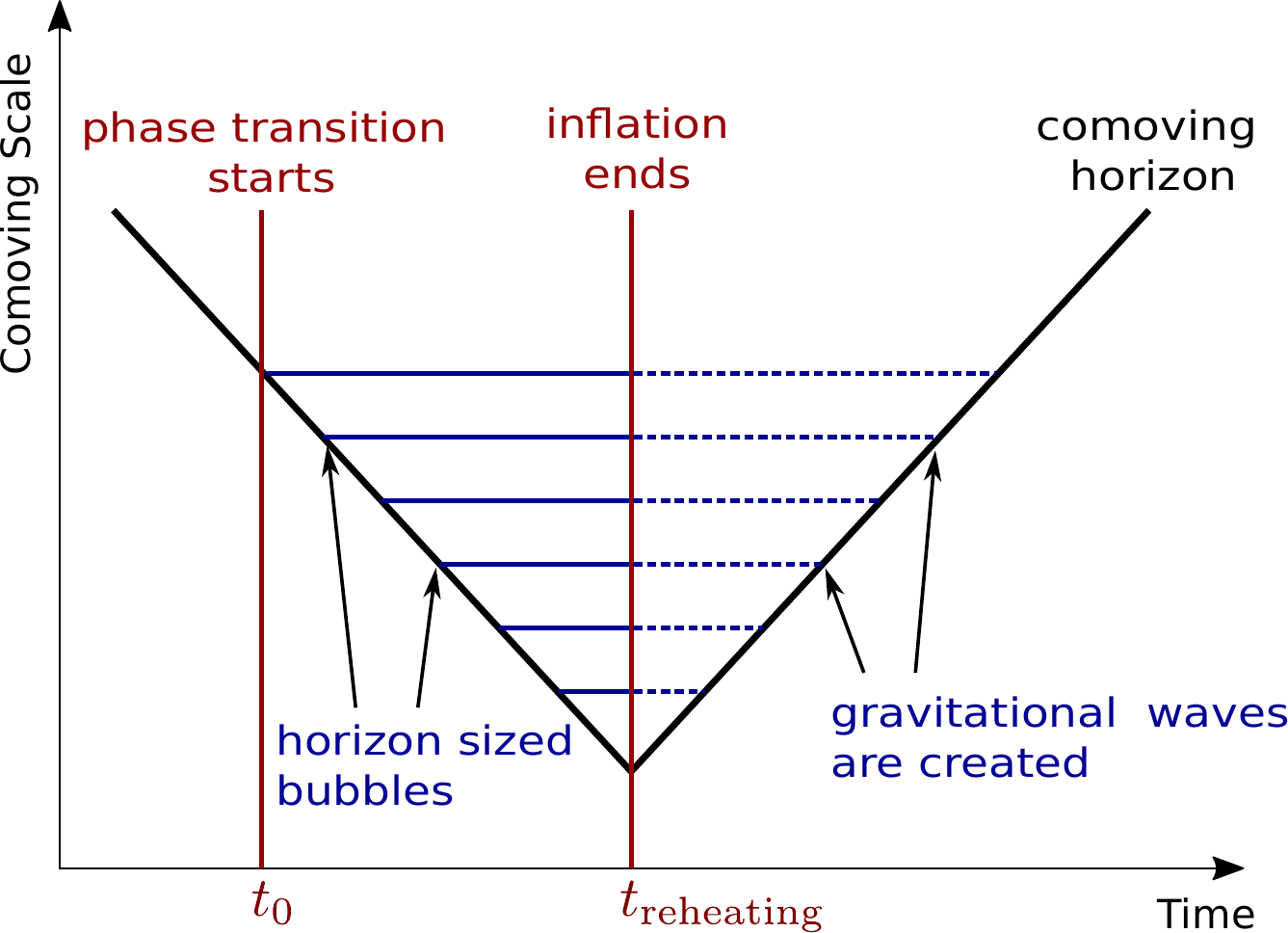}
	\caption{The process of bubble formation and GW production on comoving scales.
	Bubbles are formed equally at all times from the beginning of the phase transition and until the end of inflation.
	These bubbles quickly grow to horizon size and freeze in the case of Coleman-de Luccia tunneling, or form frozen at horizon size through Hawking-Moss tunneling.
	In either case, when inflation ends,
 	the transition completes due to the reduced expansion rate or increased temperature.
 	The inhomogeneity introduced by the bubbles remains, generating GWs upon horizon re-entry.}
	\label{fig:horizon}
\end{figure}

The process of horizon exit and re-entry is illustrated in Fig. \ref{fig:horizon}.
The bubbles are either created small and rapidly expand to horizon size, as in the case of CdL, or created exactly at horizon size, as in the case of HM.  Either way, once at the horizon, the comoving radius is completely frozen.
After the end of inflation, the phase transition completes everywhere but the imprint on the curvature
power spectrum remains. Upon horizon re-entry, the inhomogeneities produce GWs from secondary effects.
Independent of the fine details of the model, the dynamics are governed by merely three parameters: the tunneling rate per
unit volume $\Gamma/V$, the vacuum energy difference $\Delta V_{\rm PT}$ between the false and true vacuum, and the
time $t_{0}$ at which the transition commences (shortly after $\chi$ reaches the false vacuum).   With this simplified description we now turn to calculate the GW spectrum produced by such inflationary incomplete PTs and arrive at predictions for future experiments.

%%%%%%%%%%%%%%%%
\section{Scalar Curvature Spectrum}\label{sec:scalar}
We now move to calculate the scalar spectrum. To this end, we first find the energy-momentum tensor $T_{\mu\nu}$, neglecting the energy density in the bubble wall. This is justified because the ratio of energy in the interior volume over the wall energy scales as $a$ (the metric scale factor), and grows as the universe rapidly expands.     
Given the above, we omit the spatial derivatives of $\chi$ and $\phi$, which are localized in the bubble walls, and write the energy density and pressure after the PT starts, at $t\ge t_{0}$, 
\begin{eqnarray} 
\label{eq:rho}
	\rho(t,\vec{x}) &=& \frac{1}{2}\dot{\phi}^2+V_{\rm inf}(\phi(t))+[1-\theta(t-t_{\vec{x}})]\Delta V_{\rm PT}\,,
	\\
\label{eq:p}
	p(t,\vec{x}) &=& \frac{1}{2}\dot{\phi}^2-V_{\rm inf}(\phi(t))-[1-\theta(t-t_{\vec{x}})]\Delta V_{\rm PT}\,.
\end{eqnarray}
Here $t_{\vec{x}}$ is the time when the transition occurred at point $\vec{x}$, and $\theta$ is the Heaviside step function.   Using a step function is justified under the assumption of a rapid roll to the true vacuum once $\chi$ tunnels out of the false minimum.  Furthermore, the kinetic energy stored in $\chi$ around the true minimum is quickly dissipated and is therefore  neglected. 
All other components of the energy-momentum tensor can be neglected. 

The main effect of the PT on the curvature spectrum is through the change in the Hubble constant due to the shift in the vacuum energy.  We use the linearized Einstein equations in Newtonian gauge to calculate this
induced curvature perturbation to first order. To this end, we need to find the inhomogeneous part of  $T_{\mu\nu}$:
\begin{equation}
	\rho(t,\vec{x})=\bar{\rho}(t)+\delta\rho(t,\vec{x})\,,\quad \delta \rho \ll \rho\,,
\end{equation}
and similarly for $p$. The homogeneous background is taken to be
\begin{align}\label{eq:bg_rho}
	\bar{\rho}(t) = \frac{1}{2}\dot{\phi}^2+V_{\rm inf}(\phi(t))+[1-\theta(t-\mean{t_{\vec{x}}})]\Delta V_{\rm PT}\,,  \\ \label{eq:bg_p}
	\bar{p}(t) = \frac{1}{2}\dot{\phi}^2-V_{\rm inf}(\phi(t))-[1-\theta(t-\mean{t_{\vec{x}}})]\Delta V_{\rm PT}\,,
\end{align}
while the perturbations are given by
\begin{equation} \label{eq:drho}
	\delta \rho(t,\vec{x}) = \Delta V_{\rm PT} [\theta(t-\mean{t_{\vec{x}}})-\theta(t-t_{\vec{x}})]\,,
\end{equation}
and $\delta p = - \delta \rho$.
The (scalar) perturbed metric in Newtonian gauge (for a review see, e.g.~\cite{baumann2012tasi}) is,
\begin{multline}
	\mathrm{d} s^{2}=-(1+2 \Phi) \mathrm{d} t^{2}
	+a^{2}(t)(1-2 \Psi) \left( \mathrm{d} x^2 + \mathrm{d} y^2+ \mathrm{d} z^2\right)\,,
\end{multline}
and the gauge invariant comoving curvature perturbation is defined as
\begin{equation}
	\R=\Psi-\frac{H}{\bar{\rho}+\bar{p}}\delta q\,,
\end{equation}
where $\delta q$ is the scalar momentum perturbation. When specifying the energy momentum tensor in Eqs.~\eqref{eq:bg_rho}, \eqref{eq:bg_p} and \eqref{eq:drho}, we have neglected the wall energy, which is equivalent to setting $\delta q = 0$ and $\R=\Psi$. 

The only Einstein equation we will need for calculating $\R$ is
\begin{equation} \label{eq:einstein}
	\dot{\R}+H\Phi = 0\,.
\end{equation}
Using the continuity equation
\begin{equation}
	\delta p + (\bar{\rho}+\bar{p})\Phi = 0\,,
\end{equation}
we extract $\Phi$ and plug the result into Eq.~\eqref{eq:einstein},
\begin{equation} \label{eq:Rdot}
	\dot{\R} = H\frac{\delta p}{\bar{\rho}+\bar{p}}=-H\frac{\delta \rho}{\dot{\phi}^2}\,.
\end{equation}
Equation~\eqref{eq:drho} shows that $\R$ is constant for  $t<t_1=\min(t_{\vec{x}},\mean{t_{\vec{x}}})$ and for $t>t_2=\max(t_{\vec{x}},\mean{t_{\vec{x}}})$. For simplicity, and using the slow-roll approximation, we take $\dot{\phi}$ and $H$ to be constant and assume an initial flat background.  We will later relax these assumptions in order to demonstrate the sensitivity of the predicted spectrum to the inflationary dynamics.  The integrated Eq.~\eqref{eq:Rdot} then gives
\begin{equation} \label{eq:Rsol}
	\R(\vec{x}) = -\frac{H \Delta V_{\rm PT}}{\dot{\phi}^2}\left(t_{\vec{x}}-\mean{t_{\vec{x}}}\right)
	\equiv-\frac{H \Delta V_{\rm PT}}{\dot{\phi}^2} \delta t_{\vec{x}}\,.
\end{equation}

We move to calculate the scalar power spectrum $\PP_\R(k)$, defined by
\begin{equation} \label{eq:P_R_def}
      \mean{\R_{\vec{k}}\R_{\vec{k}'}}  = \delta(\vec{k}+\vec{k}')\frac{2\pi^2}{k^3}\PP_\R(k)\,.
\end{equation}
 Equation~\eqref{eq:Rsol} implies that one has to 
calculate the correlation
between the tunneling times at different points in space, $\mean{\delta t_{\xvec}\delta t_{\xvec}}$. The details of this calculation are given in Appendix~\ref{app:2_point_corr}, where we assume spherical bubbles and a constant tunneling
rate $\Gamma/V$. The calculation further assumes that bubbles nucleate frozen at horizon size, a valid assumption in the HM case, and a reasonable approximation in the CdL case.

\begin{figure}
	\centering
	\includegraphics[width=\columnwidth]{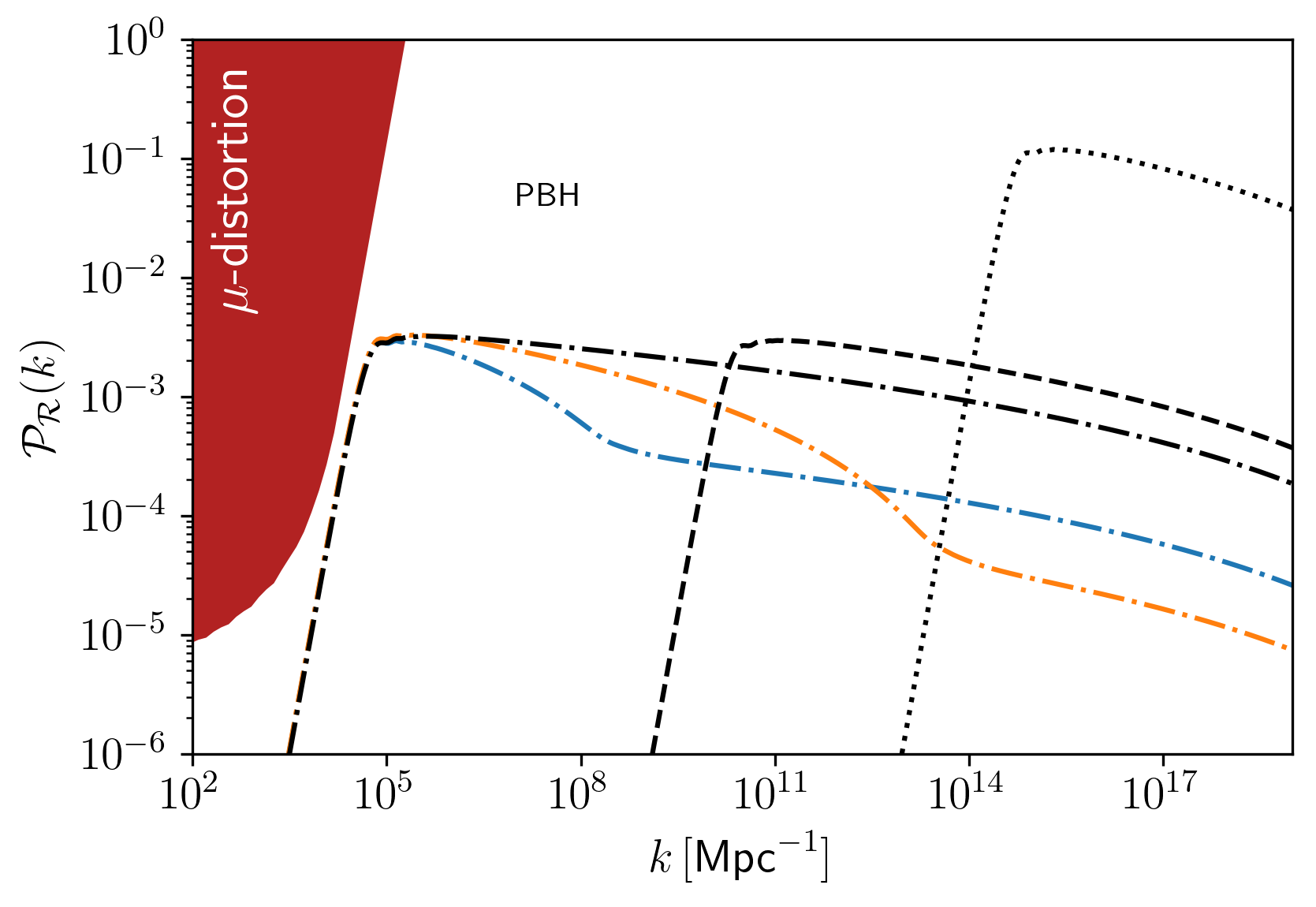}
	\caption{The scalar spectrum $\PP_{\R}$, shown for different
	choices of the parameters. The dimensionless parameter $\gamma_{\rm PT}\equiv \frac{1}{H^{4}}\frac{\Gamma}{V}\left( \frac{\Delta V_{\rm PT}}{\dot{\phi}^2} \right)^2$, which is assumed to be constant, linearly controls the amplitude.
	The momentum scale of the horizon at the beginning of the phase transition, $k_0\equiv Ha(t_0)$, acts as a minimal scale, below which the spectrum is strongly suppressed. The scale of reheating $k_{\rm re}\equiv Ha(t_{\rm reheating})$ was fixed at $4\cdot10^{22}\text{Mpc}^{-1}$.
	The {\bf black dash-dotted, dashed, and dotted} lines correspond to the parameter choices $k_0=2\cdot 10^{4},8\cdot 10^{9},2\cdot 10^{14}\,\rm{Mpc}^{-1}$ and $\gamma_{\rm PT}=5\cdot10^{-7},10^{-6},10^{-4}$, respectively.
	The {\bf blue and orange dashed-dotted} lines show the spectrum for an alternative scenario where the value of $\frac{H}{\dot{\phi}^2}$ changes by a factor of $1/10$ at time $t_{\rm drop}$.
	We define $k_{\rm drop}\equiv Ha(t_{\rm drop})$ and take its value to be $k_{\rm drop}=10^8,10^{13}\,\rm{Mpc}^{-1}$ for the blue and orange lines respectively. We further choose $k_0=2\cdot 10^{4}$ and  $\gamma_{\rm PT}=7\cdot10^{-6},2\cdot 10^{-6}$ to ensure that the peaks align, thereby demonstrating the effect of the drop in $\frac{H}{\dot{\phi}^2}$ on the spectral shape.
	The colored lines demonstrate how the spectral shape probes the dynamics of inflation over the duration of the phase transition.
	The {\bf red region} is excluded by existing bounds on CMB spectral distortions \cite{Chluba:2012we}.
	The {\bf pale red line} represents existing bounds from constraints on
	primordial black holes, taken from \cite{Carr:2020gox}. This line is
	not accurate, because known bounds assume Gaussianity.
}
	\label{fig:P_R}
\end{figure}

The scalar spectrum is shown in Fig.~\ref{fig:P_R} for three different choices of the relevant parameters. The PT commences at $t_0$ and carries on  until the end of inflation at $t_{\rm reheating}$,  continuously producing bubbles.  The effect on the primordial power spectrum therefore spans the range of momentum modes that exit the horizon at this period of time, i.e. from $k_0\equiv Ha(t_0)$ to $k_{\rm re}\equiv Ha(t_{\rm reheating})$. Since bubbles are created at a fixed rate in a universe with a shrinking co-moving Hubble sphere, the spectrum is expected to be approximately flat (varying only logarithmically).  For concreteness, throughout this work we fix $k_{\rm re}=4\cdot10^{22}\text{Mpc}^{-1}$.
Under our assumptions above of inflation with fixed slow-roll parameters, the shape and position of the peak are determined by the two $k$'s, while the amplitude further depends linearly on the dimensionless parameter $\gamma_{\rm PT} \equiv \frac{1}{H^{4}}\frac{\Gamma}{V}\left( \frac{\Delta V_{\rm PT}}{\dot{\phi}^2} \right)^2$
(see Eqs.~\eqref{eq:Rsol} and~\eqref{eq:P_R_def} as well as App.~\ref{app:2_point_corr} for details).
We find that the maximal value of the power spectrum, ${\cal P_R}$ only weakly depends on $k_0$ and is roughly ${\cal P}_{{\cal R},{\rm max}}/\gamma_{\rm PT} \approx {\cal O}(10^3)$. With the scalar spectrum calculated, we move on in Section~\ref{sec:GW} to calculate the GW spectrum it generates in the radiation-dominated era.

We emphasize that the power spectrum is sensitive not only to the spectator field which drives the PT, but also to the inflationary dynamics themselves, and its shape records the entire inflationary history from the beginning of the PT to the onset of reheating. Consequently, a measurement of the spectral shape can reveal detailed information about the dynamics of inflation.  So far, we have assumed that the slow-roll parameter and scale of inflation are constant (see Eqs.~\eqref{eq:Rdot} and~\eqref{eq:Rsol}). Relaxing this assumption strongly affects the spectrum as is demonstrated by the blue and orange dashed-dotted lines in Fig.~\ref{fig:P_R}, which assume a sudden change in the value of $\frac{H}{\dot{\phi}^2}$ occurring at time $t_{\rm drop}$. For a detailed derivation of ${\cal R}(\vec x)$ in this case, we refer the reader to App.~\ref{app:changing_phidot}.

Figure~\ref{fig:P_R} further shows constraints from the overclosure due to primordial black hole (PBH) abundance (red line), and distortions to the CMB black-body spectrum (red region). 
We note that the PBH constraint, taken from~\cite{Carr:2020gox}, assumes a Gaussian $\PP_{\R}$. 
The line shown in Fig~\ref{fig:P_R} is therefore only a rough estimation.
To derive the exact bound of PBH abundance on the spectrum, a model-specific calculation, which considers the non-Gaussian statistics of the phase transition, is required and goes beyond the scope of this paper.

%%%%%%%%%%%%%%%%
\section{Gravitational Wave Spectrum}\label{sec:GW}
After horizon re-entry, the curvature perturbations produce GWs
through second order effects. A general prescription 
for calculating GWs
induced during the radiation dominated era is derived in
\cite{Kohri_2018}, where the inhomogeneity was assumed to be Gaussian. This allows
the use of Wick's theorem to reduce the four-point correlation functions of $\R$ into
products of two-point correlations, i.e. the power spectrum $\PP_{\R}$. 
Although the phase transition spectrum is very far from Gaussianity, we have found that the result
\cite{Kohri_2018} still applies, because the "connected" part of the four-point correlation function
does not induce GWs. This point is further explained and proven in App.~\ref{app:GW}. In the following we briefly review the relevant result of~\cite{Kohri_2018}, before applying it to the spectrum derived in the previous section.

The GW energy density parameter is given by
\begin{equation} \label{eq:Omega_GW}
\frac{d\Omega_{\mathrm{GW}}}{d\log k}(\eta, k)\equiv \frac{1}{\rho_{\rm tot}}\frac{d\rho_{\rm GW}}{d\log k} =\frac{1}{24}\left(\frac{k}{a(\eta) H(\eta)}\right)^{2} \overline{\PP_{h}(\eta, k)}\, ,
\end{equation}
where $\eta$ is the conformal time and $\overline{\PP_{h}(\eta, k)}$ is the time averaged tensor spectrum, given by
\begin{equation}\begin{split} \label{eq:P_h_integral}
&\overline{\PP_{h}(\eta, k)}= 4\int_{0}^{\infty} \mathrm{d} v \int_{|1-v|}^{1+v} \mathrm{d}u  \\
&\left(\frac{4 v^{2}-\left(1+v^{2}-u^{2}\right)^2}{4 v u}\right)^{2}
\overline{\tilde{I}^2(v, u, k\eta)} \PP_{\R}(k v) \PP_{\R}(k u)\,.
\end{split}\end{equation}
Here the quantity $\overline{\tilde{I}^2(v, u, k\eta)}$ is defined in Eq.~\eqref{eq:I_radiation}.

\begin{figure}[t]
	\centering
	\includegraphics[width=\columnwidth]{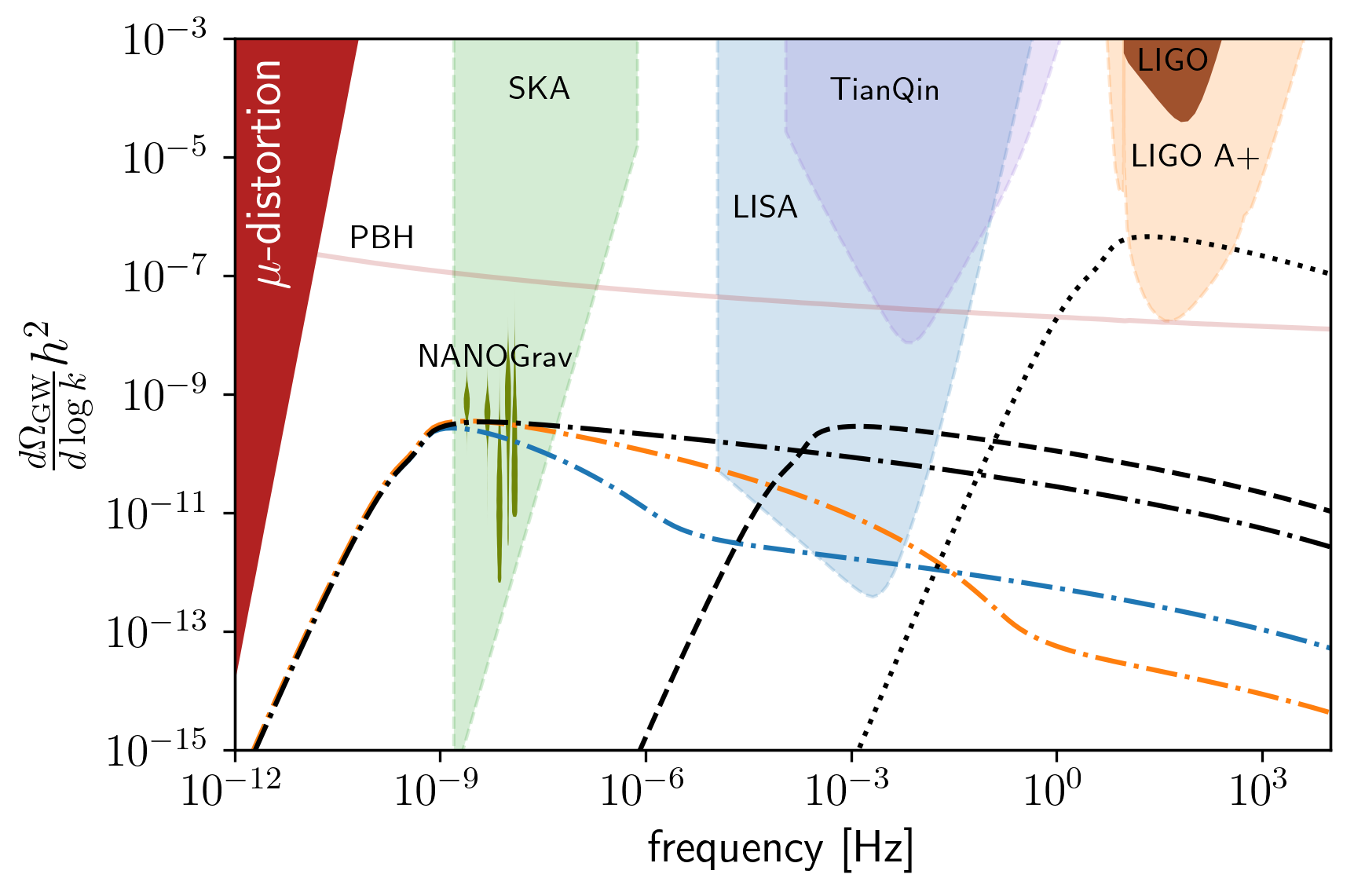}
	\caption{The GW energy density induced by the scalar spectra shown in Fig.~\ref{fig:P_R}, multiplied by the square of the scaling factor of the Hubble expansion rate, $h$. 
The amplitude is quadratically dependent on the dimensionless parameter $\gamma_{\rm PT}\equiv \frac{1}{H^{4}}\frac{\Gamma}{V}\left( \frac{\Delta V_{\rm PT}}{\dot{\phi}^2} \right)^2$, which is assumed to be constant.
	The momentum scale of the horizon at the beginning of the phase transition, $k_0\equiv Ha(t_0)$, acts as a minimal scale, below which the spectrum is strongly suppressed. The scale of reheating $k_{\rm re}\equiv Ha(t_{\rm reheating})$ was fixed at $4\cdot10^{22}\text{Mpc}^{-1}$.
	The {\bf black dash-dotted, dashed, and dotted} lines correspond to the parameter choices $k_0=2\cdot 10^{4},8\cdot 10^{9},2\cdot 10^{14}\,\rm{Mpc}^{-1}$ and $\gamma_{\rm PT}=5\cdot10^{-7},10^{-6},10^{-4}$, respectively.
	The {\bf dashed-dotted blue and orange} lines show the spectrum for an alternative scenario where the value of $\frac{H}{\dot{\phi}^2}$ changes by a factor of $1/10$ at time $t_{\rm drop}$.
	We define $k_{\rm drop}\equiv Ha(t_{\rm drop})$ and take its value to be $k_{\rm drop}=10^8,10^{13}\,\rm{Mpc}^{-1}$ for the blue and orange lines respectively. We further choose $k_0=2\cdot 10^{4}$ and  $\gamma_{\rm PT}=7\cdot10^{-6},2\cdot 10^{-6}$ to ensure that the peaks align, thereby demonstrating the effect of the drop in $\frac{H}{\dot{\phi}^2}$ on the spectral shape.
	Current	constraints and future detector sensitivity regions are shown with {\bf solid}, and {\bf semi-transparent} colored regions respectively.
	The detector sensitivity curves for SKA \cite{5136190}, LISA
	\cite{amaroseoane2017laser} \cite{Sathyaprakash:2009xs},
	TianQin \cite{TianQin:2015yph}, Ligo \cite{KAGRA:2013rdx}, and
	Ligo A+, are taken from~\cite{Moore:2014lga}.
	The {\bf green violin} plots represent the free-spectrum fit to the
	NANOGrav data \cite{NANOGrav:2020bcs}\cite{Ratzinger:2020koh}.
	The {\bf red region} and {\bf pale red line} are CMB distortion and primodrial black hole constraints derived from those in Fig.~\ref{fig:P_R}}.
	\label{fig:GW}
\end{figure}

$d\Omega_{\rm GW}/d\log k$ approaches a constant value during radiation domination because the GWs redshift like radiation. The density during radiation domination can thus be related to the density today through
\begin{equation} \label{eq:Omega_today}
	\frac{d\Omega_{GW}(\eta_0,k)}{d\log k}=\Omega_r(\eta_0)\frac{\Omega_{GW}(\eta_c,k)}{d\log k}\,,
\end{equation}
where $d\Omega_{GW}(\eta_c,k)/d\log k$ is the constant value reached during the radiation dominated era, and $\Omega_r(\eta_0)\approx 10^{-4}$ is the energy fraction of radiation today.

Integrating Eq.~\eqref{eq:P_h_integral}
numerically, we obtain the GW spectrum shown in Fig.~\ref{fig:GW} for the parameters discussed in Sec.~\ref{sec:scalar}. 
Since the scalar spectrum
is almost scale-invariant over a wide range of momenta, the induced abundance of 
GWs can be approximated by the analytical
result given in~\cite{Kohri_2018} for a scale invariant case, $d\Omega_{GW}(\eta_c,k)/d\log k \simeq 0.8\PP_{\R}^2$.
Applying this approximation along with~
Eq.~\eqref{eq:Omega_today}, we find that much like the power spectrum, the peak of the GW energy density today can be
estimated directly from the model parameters,
\begin{equation}
\frac{d\Omega_{GW}(\eta_0,k)}{d\log k}\Bigg|_{\rm peak}\simeq {\cal O}(100) \gamma_{\rm PT} ^2.
\end{equation} 
In addition to the predicted lines, Fig.~\ref{fig:GW} shows the corresponding constraints from Fig.~\ref{fig:P_R}, as well as constraints from LIGO~\cite{KAGRA:2013rdx} and expected sensitivity from various future GW detectors.  See caption for details.

%%%%%%%%%%%%%%%%%%
\section{Discussion} \label{sec:discussion}
An incomplete phase transition occurring in the late stages of
inflation may generate observable gravitational waves, summarized in Fig.~\ref{fig:GW}.
Since the transition occurs over a long period of time, the resulting spectrum is very wide.
The unique shape allows it to be detected by future experiments and to be distinguished from other scenarios, which typically predict a specific peak frequency.  Furthermore, as we showed, the resulting spectrum depends on the details of the inflationary dynamics and a corresponding measurement would probe the slow-roll parameters at small scales.  

The model studied in this paper may produce a significant amount of primordial
black holes. Calculating the primordial black hole abundance requires a
detailed study of the gravitational collapse which we leave for future work.

\mysections{Acknowledgments}
We thank Nadav Outmezguine for useful discussions.  The work of 
TV is supported by the Israel Science Foundation (grant No. 1862/21), by the Binational Science Foundation (grant No. 2020220) and by the European Research Council (ERC) under the EU Horizon 2020 Programme (ERC-CoG-2015 - Proposal n. 682676 LDMThExp).  MG
is supported in part by Israel Science Foundation under Grant No. 1302/19. MG is also supported in part
by the US-Israeli BSF grant 2018236
%%%%%%%%%%%%%%%%%%%%%

\appendix
%TC:ignore
%%%%%%%%%%%%%
\section{The 2-point correlation function of the tunneling time} \label{app:2_point_corr}
Here we calculate the correlation function of $\delta t_{\vec{x}}$ defined in Eq.~\eqref{eq:Rsol},
\begin{equation} \label{eq:t_to_dt}
    \langle \delta t_{\vec{x}} \delta t_{\vec{x}'} \rangle =
    \langle t_{\vec{x}} t_{\vec{x}'} \rangle - \langle t_{\vec{x}}\rangle^2
\end{equation}
as a function of $r=|\vec{x}-\vec{x}'|$.
For simplicity, we will ignore the end of inflation for now, and calculate the correlation as if inflation goes on forever.
In the next section, we will show how the calculation has to be modified in order to account for the end of inflation.

After deriving the detailed correlation function, we find a simple approximate expression valid under the assumption of slow tunneling rate,  $\Gamma/V\ll H^4$.
This simplification will come handy in the next section where we compute the
effect of non-Gaussianity.

\subsection{The correlation without ending inflation}
To compute the correlation function, we first need to find the probability distribution for the decay of the false vacuum. For a general phase transition, the probability to find a given point in space in the false vacuum by the time $\bt$ is given by \cite{Turner:1992tz}
\begin{equation} \label{decay probabilty}
 p(t_{\xvec} > \bt) =
	 e^{-I(\bt)}
\end{equation}
with,
\begin{equation}\label{eq:I_def}
	  I(\bt)=\frac{4\pi}{3} \int_{t_0}^{\bt} d\bt' \frac{\Gamma}{V} (\bt')       a^3(\bt') r^3(\bt,\bt')\,.
\end{equation}
Here $I(\bt)$ is the ratio of the space volume inside the bubbles over the entire volume at time $\bt$, without accounting for the overlaps.  The latter are automatically taken care of by the exponent in Eq.~\eqref{decay probabilty}. Note that by  definition, Eq.~\eqref{decay probabilty} is the complementary cumulative probability. In Eq.~\eqref{eq:I_def}, $r(\bt,\bt')$ is the comoving radius of a bubble that was created at  time $\bt'$ and measured at time $\bt$. $\frac{\Gamma}{V}(\bt')$ is the tunneling rate per unit volume at time $\bt'$, and $t_0$ the time when the phase transition commences. We will take $t_0=0$ to simplify the expressions.

For the toy models studied in this work, ${\Gamma}/{V}$ is constant.  We further assume that at formation, the bubble radius coincides with the Hubble radius and thus,
\begin{equation}
r(\bt,\bt')=({a_0e^{H\bt'}H})^{-1}\equiv r_\mathcal{H}(\bt'),
\end{equation}
where $a_0$ is the scale factor at $t=0$. We stress that this expression is exact for the case of the HM transition, but is only  approximate for the CdL transition. In the latter case, we neglect the time it takes the bubble to grow to horizon-size. 
Plugging the radius back into Eq.~\eqref{eq:I_def} gives the exponential decay
\begin{align} \label{eq:exp_probability}
	p(t_{\vec{x}} > \bt) &= 
	 e^{-\bt/\tau}, 
\end{align}
with $\tau$ being the mean lifetime of the false vacuum at any given point
\begin{align} \label{eq:tau_def}
%\frac{1}{\tau}
\tau
	 & =
	 \left [\frac{4\pi}{3} 
	 %\right )
	 \left ( \frac{1}{H^3}
	 \right )
	 \left (
	 \frac{\Gamma}{V}
	 \right )
	 \right ]^{-1}.
\end{align}
which agrees with a more direct calculation.

Now the second term in Eq.~\eqref{eq:t_to_dt} can be extracted by taking the mean of the probability at Eq.~\eqref{eq:exp_probability}:
\begin{equation} \label{eq:t_mean}
    \langle t_{\vec{x}}\rangle = %\frac{3H^3}{4\pi}\left(\frac{\Gamma}{V}\right)^{-1} .  
    \tau.
    \end{equation}
The remaining part of this section is focused on computing the covariance $\mean{t_{\vec{x}}  t_{\vec{x}'}}$. We use the law of total expectation 
\begin{equation}
	\langle t_{\vec{x}} t_{\vec{x}'}\rangle = \langle t_{\vec{x}} E(t_{\vec{x}'}|t_{\vec{x}})\rangle\,,
\end{equation}
where $E(\cdot)$ is the conditional expectation value.
 Applying this to the exponential probability distribution described by Eq.~\eqref{eq:exp_probability} gives
\begin{equation}
 \label{eq:total_expectation}
    \langle  t_{\vec{x}}  t_{\vec{x}'} \rangle 
    = 
    %\frac{4\pi}{3 H^3} \frac{\Gamma}{V} 
    \frac{1}{\tau}
    \int_0^{\infty} 
    dt_{\vec{x}}
    \, t_{\vec{x}}
    \,
    %e^{-\frac{4\pi}{3 H^3} \frac{\Gamma}{V}  t_{\vec{x}}}
    e^{-t_{\vec{x}}/\tau}
    E(t_{\vec{x}'}|t_{\vec{x}})\,.
\end{equation}

With the above, we are left with computing the conditional expectation value, $E(t_{\vec{x}'}|t_{\vec{x}})$. 
This describes the expectation of $t_{\xvec'}$ assuming we know $t_{\xvec}$.
Given the cumulative probability distribution in the domain of $(0, +\infty)$, the expectation is given by
\begin{align}
\label{eq:cdf-to-expectation}
    E(t_{\xvec'} | t_{\xvec})
    & = 
    \int_0^\infty d \bt
    \; 
    p(t_{\xvec'} > \bt | t_{\xvec}).
\end{align}
To compute Eq.~\eqref{eq:total_expectation}, we separate the integral over $t_{\xvec}$ into two intervals, $t_{\xvec} \in [0, t_{\rm s})$ and $t_{\xvec} \in [t_{\rm s},\infty)$, where $t_{\rm s}$  is the separation time, defined to be the moment when the distance between $\xvec$ and $\xvec'$ becomes larger than twice the Hubble radius, \textit{i.e.} $r = 2r_\mathcal{H}(t_{\rm s})$. One finds,
\begin{equation} \label{seperation time}
    t_{\rm s} = \frac{1}{H} \log\frac{2}{a_0Hr}.
\end{equation}
When the points are more than two radii apart, they can no longer be contained inside any single bubble, and thus $t_{\rm s}$ represents the moment at which  these two points became independent.
\subsubsection{first interval: $0 < t_{\xvec}<t_{\rm s}$}
Let us first consider the case where the false vacuum at $\xvec$ decays before the two points are causally disconnected.  
In principle a bubble that contains $\xvec'$ may form at any time $\bt$. To evaluate the integral in Eq.~\eqref{eq:cdf-to-expectation}, we further break it into two intervals, $0\le \bt < t_{\xvec}$ and $\bt\ge t_{\xvec}$. We begin with the first interval.

 In the case where $\bt<t_{\xvec}$, the bubble cannot contain $\xvec$. We therefore have to modify Eq.~\eqref{eq:I_def} in order exclude from $I$ the contribution of bubbles which form at $\bt<t_{\xvec}$ and include $\xvec$.
The centers of all possible bubbles that could form at $\bt$ containing $\xvec'$ form a ball of radius $r_\mathcal{H}(\bt)$ centered at $\xvec'$. Similarly, the centers of bubbles containing $\xvec$ form a Hubble ball of the same radius centered at $\xvec$. Therefore, if we randomly draw a bubble that contains $\xvec'$, the probability this bubble doesn't contain $\xvec$ is given by the fraction of the Hubble ball centered at $\xvec'$ that doesn't overlap with the ball centered at $\xvec$:
\begin{align} \label{eq:bubble_nonoverlap}
    f_V(\bt)
    & =
    \frac{V_\mathcal{H}(\bt) - V_{\rm O}(\bt)}{V_\mathcal{H}(\bt)}
    = 
    \frac{3r}{4r_\mathcal{H}(\bt)} - 
    \frac{r^3}{16 r_\mathcal{H}^3(\bt)},
\end{align}
where the Hubble volume is $V_\mathcal{H} = 4\pi r_\mathcal{H}(\bt)^3/3$ and the overlapping volume is $V_{\rm O}=\pi(4r_\mathcal{H}(\bt)+r)(2r_\mathcal{H}(\bt)-r)^2/12$. 
Now, excluding from \eqref{eq:I_def} the bubbles that include $\xvec$ means replacing $I(\bt)$ with 
\begin{align}
\label{eq:J_def}
        J(\bt)
        & = \frac{4\pi}{3} \int_{0}^{\bt} d\bt' \frac{\Gamma}{V} a^3(\bt') r_\mathcal{H}^3(\bt') f_V(\bt')\cr
        & = 
        \frac{1}{\tau H}
        \left (
        \frac{3 r}{4r_\mathcal{H}(\bt')} 
        - \frac{r^3}{48r_\mathcal{H}^3(\bt')}
        \right )
        \bigg |^{\bt'=\bt}_{\bt' = 0}.
\end{align}
For $\bt<t_{\xvec}$, the probability is thus given by,
\begin{equation}\label{eq:probability_early_small_r}
    p(t_{\xvec'}>\bt|t_{\xvec}) = 
    e^{-J(\bt)}\,.
\end{equation}

We move on to the interval $\bt\ge t_{\xvec}$. At $\bt=t_{\xvec}$, we know a bubble forms around $\xvec$. To compute the probability this bubble does not include $\xvec'$ let us first map out the allowed region for the bubble center. To contain $\xvec$ the bubble center can be placed anywhere in a sphere that is centered at $\xvec$ with a radius of $r_\mathcal{H}(t_{\xvec})$. To avoid containing $\xvec'$, the center of the bubble cannot be anywhere inside a sphere  centered at $\xvec'$ with  radius $r_\mathcal{H}(t_{\xvec'})$.
Therefore, if we randomly place a bubble such that it contains $\xvec$, the probability it does not contain $\xvec'$ is given by the non-overlapping fraction $f_V(t_{\xvec})$.
As a result, the probability that $\xvec'$ will remain in the false vacuum when the bubble around $\xvec$ is formed, is given by
\begin{equation}\label{eq:p-at-tx}
p(t_{\xvec'}>t_{\xvec}|t_{\xvec}) =e^{-J(t_{\xvec})}
f_V(t_{\xvec}).
\end{equation}
We note that the discontinuity of $p(t_{\xvec'} > \bt | t_{\xvec})$ at $\bt = t_{\xvec}$ (see Eqs.~\eqref{eq:probability_early_small_r} and~\eqref{eq:p-at-tx}), is due to the instantaneous formation of a bubble at $\xvec$. 

The known value of $t_{\xvec}$ does not pose any constraints on bubble formation after $t_{\xvec}$, so for $\bt>t_{\xvec}$ the probability will decay exponentially with the same rate as in Eq.~\eqref{eq:exp_probability}:
\begin{equation} \label{eq:probability_late_small_r}
    p(t_{\xvec'}>\bt|t_{\xvec}) = 
    e^{-J(t_{\xvec})}
    f_V(t_{\xvec})    
    e^{-(\bt - t_{\xvec})/\tau}.
\end{equation}
Finally, putting Eq.~\eqref{eq:probability_early_small_r} and Eq.~\eqref{eq:probability_late_small_r} into Eq.~\eqref{eq:cdf-to-expectation} we obtain the conditional expectation
\begin{multline} \label{eq:conditional E small r}
    E(t_{\xvec'}|t_{\xvec})=
    \int_0^{t_{\xvec}} e^{-J(\bt)} d\bt +		
    \tau \,
    e^{-J(t_{\xvec})}
    f_V(t_{\xvec})
\end{multline}

\subsubsection{second interval: $t_{\xvec}\ge t_{\rm s}$}
This corresponds to the scenario where the false vacuum at $\xvec$ decays after the two bubbles are causally disconnected.  
For $\bt<t_{\rm s}$ the probability is given by Eq.~\eqref{eq:probability_early_small_r} as before. At $t_{\rm s}$ the two points become separated and the decay becomes independent. Therefore, $t_{\xvec}$ is irrelevant to the conditional probability. The cumulative distribution function is given  by 
\begin{equation} \label{eq:probability:all times:large d}
    p(t_{\xvec'}>\bt|t_{\xvec}) = 
    \begin{cases}
    e^{-J(\bt)},& \bt < t_{\rm s} \\
    e^{-J(t_{s})}
    e^{-(\bt-t_{s})/\tau} ,      & \bt \ge t_{\rm s}
\end{cases}
\end{equation}
which means the conditional expectation is
\begin{equation} \label{eq:conditional E large r}
    E(t_{\xvec'}|t_{\xvec}) =
    \int_0^{t_{\rm s}} e^{-J(\bt)} d\bt +
    \tau \,
    e^{-J(t_{\rm s})}
    .
\end{equation}
Since Eq.~\eqref{eq:conditional E large r} is independent of $t_{\xvec}$, it can be integrated analytically when plugged back into Eq.~\eqref{eq:total_expectation}. 

Putting both intervals back into Eq.~\eqref{eq:total_expectation} we reach the final expression for the two-point expectation value:
\begin{equation}\begin{split} \label{eq:corr_final}
\langle  t_{\vec{x}}  t_{\vec{x}'} \rangle 
     =& \frac{1}{\tau}  \int_0^{t_{s}} \,dt_{\vec{x}}\, t_{\vec{x}} \, e^{-t_{\vec{x}}/\tau} E(t_{\vec{x}'}|t_{\vec{x}})+ \\
     &\left[   \int_0^{t_{\rm s}} e^{-J(\bt)} d\bt +    e^{-J(t_{\rm s})}\, \tau\,
     \right] \, 
     %\tau \, 
     e^{-t_{s}/\tau}
     \left(
     %\frac{t_{s}}{\tau}+1
     t_{\rm s} + \tau
     \right) 
\end{split}\end{equation}
% \end{widetext}
where in the first term we plug in Eq.~\eqref{eq:conditional E small r} and integrate numerically. This result is only valid for $t_{\rm s}>0$. If $t_{\rm s}\le0$, the points were separated before the phase transition started, so they decay independently and $\mean{\delta t_{\vec{x}}  \delta t_{\vec{x}'}}=0$.
%%%

\subsection{Adding the end time of inflation}
Now, we have to repeat the above calculation, taking into account the end of inflation at $t_{\rm e}=t_{\rm reheating}$, when reheating starts.
As discussed in the main text, we assume that  all of the volume of space which remained in the false vacuum
will move to the true vacuum immediately at $t_{\rm e}$, which means the probability
distribution Eq.~\eqref{eq:exp_probability} has to be replaced with
\begin{equation} \label{eq:exp_probability_fin_time}
	p(t_{\vec{x}} > \bt) = 
\begin{cases}
	 e^{-\bt/\tau}, &\bt<t_{\rm e} \\
	 0, &\bt\ge t_{\rm e}
 \end{cases}
\end{equation}
and the corresponding expectation value Eq.~\eqref{eq:t_mean} is replaced with
\begin{equation} \label{eq:t_mean_with_end}
    \langle t_{\vec{x}}\rangle = 
    \tau
    \left( 1-e^{-t_{\rm e}/\tau} 
    \right) .
\end{equation}
Note that having an end of inflation is similar to putting a regulator to $\langle t_{\vec{x}}\rangle$. Without $t_{\rm e}$ the expected transition time $\tau$ diverges in the slow tunneling limit ($\Gamma/(VH^4) \rightarrow 0$). With the new probability distribution given in Eq.~\eqref{eq:exp_probability_fin_time}, the expectation approaches $t_{\rm e}$ in the limit of slow tunneling.

The law of total expectation can be applied in the same manner as above, but now the integral in Eq.~\eqref{eq:total_expectation} ends at $t_{\rm e}$:
\begin{equation} \label{eq:total_expectation_fin_time}
	\begin{split}
    \langle  t_{\vec{x}}  t_{\vec{x}'} \rangle 
   = \frac{1}{\tau} 
   \int_0^{t_{\rm e}} dt_{\vec{x}}
   \,t_{\vec{x}} \,
    e^{-t_{\vec{x}}/\tau} E(t_{\vec{x}'}|t_{\vec{x}}) 
    +
    t_{\rm e}\,  e^{-t_{\rm e}/\tau}\, E(t_{\xvec'}|t_{\rm e})
	\end{split}
\end{equation}
where  the second term accounts for the finite probability that $\xvec$
does not tunnel until the end of inflation, $P(t_{\xvec}=t_{\rm e})=e^{-t_{e}/\tau}$.

The derivation of the conditional expectation is the same as in the previous section, except for the fact 
that the integrals end at $t_{\rm s}$ instead of $\infty$. We are only interested in
scales which exited the horizon before the end of inflation, which means $t_{\rm s}<t_{\rm e}$.
Under this assumption, we split the conditional expectation to two cases as above, 
$t_{\xvec}<t_{\rm s}$ and $t_{\rm s}< t_{\xvec} \leq t_{\rm e}$. In the first case, Eq.~\eqref{eq:probability_early_small_r} and Eq.~\eqref{eq:probability_late_small_r} remain unchanged, but the integral over $\bt$ ends at $t_{\rm e}$, which means Eq.~\eqref{eq:conditional E small r} 
has to be replaced with
\begin{multline} \label{eq:E_conditional_early_fin_time}
        E(t_{\xvec'}|t_{\xvec})=
    \int_0^{t_{\xvec}} e^{-J(\bt)} d\bt 		\\
    + e^{-J(t_{\xvec})}
    f_V(t_{\xvec})
    \,\tau \, 
    \left( 1-e^{-(t_{\rm e}-t_{\xvec})/\tau} \right) .
\end{multline}
Similarly, Eq.~\eqref{eq:conditional E large r} has to be replaced with
\begin{multline}  \label{eq:E_conditional_late_fin_time}
    E(t_{\xvec'}|t_{\xvec})=
    \int_0^{t_{\rm s}} e^{-J(\bt)} d\bt 		\\
    + e^{-J(t_{\rm s})}
    \, \tau\,
    \left( 1-e^{-(t_{\rm e}-t_{\rm s})/\tau} \right) .
\end{multline}
Eq.~\eqref{eq:E_conditional_early_fin_time} and
Eq.~\eqref{eq:E_conditional_late_fin_time} can be plugged into
Eq.~\eqref{eq:total_expectation_fin_time} and integrated numerically to get the
required correlation $\mean{t_{\xvec}t_{\xvec'}}$.

\subsection{The small $\Gamma/V$ limit} \label{app:small_Gamma_2point}
The above derivation allows the 2-point correlation to be calculated without assuming anything about $\Gamma$. Here, we derive a simpler closed-form formula by assuming $\Gamma/(VH^3) \ll t_{\rm e}^{-1}$. 
This assumption makes the resulting power spectrum linear in $\Gamma$, and the
approximation will be used in the next section to show that non-Gaussianity can
be ignored when calculating the GW spectrum.

Once again, we denote by $t_{\rm s}$ the moment at which the points $\xvec$ and $\xvec'$ become causally separated, and consider the case in which  the two points we are looking at were separated before the end of inflation, $t_{\rm s}<t_{\rm e}$. Since the rate of bubble nucleation is very small, the probability that more than a single bubble formed around any of the two points before the separation time  $t_{\rm s}$, can be neglected. The correlation of $\delta t_{\xvec}$ can therefore be written as
\begin{equation} \begin{split} \label{eq:corr_2_point_small_Gamma}
	\mean{\delta t_{\xvec}\delta t_{\xvec'}} =&E(\delta t_{\xvec}\delta t_{\xvec'}|
 \substack{\text{bubble with both} \\ \text{points before $t_{\rm s}$}})
 P(\substack{\text{bubble with both} \\ \text{points before $t_{\rm s}$}})
  \\
 +&E(\delta t_{\xvec}\delta t_{\xvec'}|
 \substack{\text{bubble with} \\ \text{only $\xvec$ before $t_{\rm s}$}})
 P(\substack{\text{bubble with} \\ \text{only $\xvec$ before $t_{\rm s}$}}) \\
  +&E(\delta t_{\xvec}\delta t_{\xvec'}|
 \substack{\text{bubble with} \\ \text{only $\xvec'$ before $t_{\rm s}$}})
 P(\substack{\text{bubble with} \\ \text{only $\xvec'$ before $t_{\rm s}$}}) \\
+&  E(\delta t_{\xvec}\delta t_{\xvec'}|
 \substack{\text{no bubble} \\ \text{before $t_{\rm s}$}})
 P(\substack{\text{no bubble} \\ \text{before $t_{\rm s}$}})
\end{split} \end{equation} 
We now show that in the $\Gamma/V\ll H^4$ limit, the first term in  Eq.~\eqref{eq:corr_2_point_small_Gamma} becomes linear in $\Gamma/V$, while the other three terms are of order $\left(\Gamma/V\right)^2$.

Consider first the last term.  Since the points become independent after $t_{\rm s}$, the expectation value in the that term 
can be written as
\begin{equation}
E(\delta t_{\xvec}\delta t_{\xvec'}| \substack{\text{no bubble} \\ \text{before $t_{\rm s}$}}) = 
E(\delta t_{\xvec}| t_{\xvec}>t_{\rm s})^2.
\end{equation} 
This expectation can be calculated by integrating the probability distribution given in
Eq.~\eqref{eq:exp_probability_fin_time} from $t_{\rm s}$ to $t_{\rm e}$,
\begin{align}
\label{eq:delta_t_after_ts} 
	E(\delta t_{\xvec}|t_{\xvec}>t_{\rm s}) 
	&= 
	t_{\rm s}+ \tau  \,
	e^{-t_{\rm e}/\tau}
	\left( 1-e^{t_{\rm s}/\tau} \right)  \cr
	&= \frac{t_{\rm s}}{\tau} \left ( t_{\rm e} - t_{\rm s}/2 \right )
	+ \mathcal O ((t_{\rm s}/\tau)^2).
\end{align}
Together with Eq.~\eqref{eq:tau_def}, this result shows the last term in Eq.~\eqref{eq:corr_2_point_small_Gamma} is at least of order
$(\Gamma/V)^2$. Similarly, the expectation value in the second and third terms can be factorized, e.g.,
\begin{equation} \begin{split}
&E(\delta t_{\xvec}\delta t_{\xvec'}|
 \substack{\text{bubble with} \\ \text{only $\xvec'$ before $t_{\rm s}$}}) = \\
&E(\delta t_{\xvec'}|
 \substack{\text{bubble with} \\ \text{only $\xvec'$ before $t_{\rm s}$}})
E(\delta t_{\xvec}|
t_{\xvec}>t_{\rm s}),
\end{split}\end{equation} 
and Eq.~\eqref{eq:delta_t_after_ts} implies that the second factor is linear in $\Gamma/V$. 
Since the first factor is regular in $\Gamma/V$, the whole expression is at least linear in $\Gamma/V$.
Given that the probabilities of forming a bubble are also  linear in $\Gamma/V$, the second and
third terms of Eq.~\eqref{eq:corr_2_point_small_Gamma} must be at least of order $(\Gamma/V)^2$.

Finally, we are ready to calculate the dominant term: the contribution to the correlation
from a single bubble forming around both points. This contribution can be rewritten as
\begin{equation} \begin{split} \label{eq:corr_2_point_1_bubble}
&\int_{0}^{t_{\rm s}} E(\delta t_{\vec{x}} \delta t_{\vec{x}'}|\substack{\text{
bubble with } \\ \text{both points at t}})dP\left( \substack{\text{bubble with} \\ \text{both points at t}} \right) 	\\
&= \int_{0}^{t_{\rm s}} \, \left( t-\mean{t_{\xvec}} \right) ^2
\frac\Gamma V  V_{\text{overlap}}(t)dt
\end{split}\end{equation} 
where $dP\propto \Gamma dt$ is the probability of forming a bubble in the infinitesimal time interval $dt$, and $V_{\rm overlap}(t)$ is the physical volume of overlap between two Hubble
spheres, each centered at $\xvec$ and $\xvec'$.
The expectation value $\mean{t_{\xvec}}$ is given by Eq.~\eqref{eq:t_mean_with_end},
but since we are calculating only the first order in $\Gamma$ we can take
$\mean{t_{\xvec}}\simeq t_{\rm e}$.
From this point, it is straightforward to write
$V_{\rm overlap}$ explicitly and integrate Eq.~\eqref{eq:corr_2_point_1_bubble} directly. Instead, we
introduce a trick that will be useful later. The overlap
volume between two spheres can be written as an integral
over a $\delta$ function:
\begin{equation}
	V_{\text{overlap}}(t)=a^3(t)\int_{r_\mathcal{H}}^{} d^3y_1 \int_{r_\mathcal{H}}^{} d^3y_2 \,
	\delta\left( \vec{y}_1-\vec{y}_2+\vec{r} \right) 
\end{equation} 
where both integrals are over a sphere of radius $r_\mathcal{H}$ centered at the origin, and $\vec{r}=\xvec_1-\xvec_2$ is the separation between the centers of the overlapping spheres.
This expression automatically vanishes for $t>t_{\rm s}$, so the upper limit of the
integral in Eq.~\eqref{eq:corr_2_point_1_bubble} can be replaced with $\infty$. After doing so, the only dependence
of Eq.~\eqref{eq:corr_2_point_1_bubble} on $\xvec_1$ and $\xvec_2$ is through $\vec{r}$ in the $\delta$ function.
That makes taking the Fourier transform trivial,
\begin{align}
    \label{eq:2_point_approx_final}
    \mean{\delta t_{\vec{k}_1} \delta t_{\vec{k}_2}} 
    &=	\delta( \vec{k}_1+\vec{k}_2 )  \left (\frac{12\pi}{k^2} \right )
    \\
	& ~~~ \times
	\frac{1}{\tau H} 
	\int_{0}^{\infty}  \frac{dt}{a(t)} \left( t-\mean{t_{\xvec}} \right) ^2
	j_1^2(kr_\mathcal{H}) \nonumber
\end{align}
where $j_1(x)=\frac{1}{x}\left( \frac{\sin x}{x}-\cos x \right) $ is the first spherical Bessel function.
Using $a(t)=a_0 e^{Ht}$ and $r_\mathcal{H}(t)=\left( Ha(t) \right) ^{-1}$ the integral
can be evaluated numerically, and by comparing with Eq.~\eqref{eq:Rsol} and
Eq.~\eqref{eq:P_R_def} we can extract the spectrum $\mathcal{P}_{\R}$.
This result is in agreement with the full calculation of the previous section
in the $\Gamma/V\to 0$ limit.

\section{Alleviating the assumption of constant $\dot{\phi}$ and $H$} \label{app:changing_phidot}
In deriving Eq.~\eqref{eq:Rsol} from Eq.~\eqref{eq:Rdot}, we assumed
$H/\dot{\phi}^2$ to be constant, which resulted with $\R$ being a linear function of $t_{\xvec}$. This result meant we only have to calculate correlations of the tunneling times $t_{\xvec}$, and then convert the final results to correlations of $\R$ with the use of Eq.~\eqref{eq:Rsol}. We will now discuss how the spectrum can be approximated without this simplification.

$\R$ is obtained by integrating Eq.~\eqref{eq:Rdot} over $t$, with an initial condition of $\R=0$. The theta functions in Eq.~\eqref{eq:drho} ensure the integrand vanishes unless t is between $\mean{t_{\xvec}}$ and $t_{\xvec}$, allowing us to write the integral as
\begin{equation} \label{eq:R_integral}
    \R(t_{\xvec}) = -\Delta V_{\rm PT}\int_{\mean{t_{\xvec}}}^{t_{\xvec}}
    \frac{H}{\dot{\phi}^2}dt\,.
\end{equation}
Eq.~\eqref{eq:Rsol} can be recovered from this more general result by assuming the integrand is constant.

As we have shown in Section \ref{app:small_Gamma_2point}, in the limit of small $\Gamma/V$ we only have to consider the contribution of a single bubble to the correlation function, given by Eq.~\eqref{eq:corr_2_point_1_bubble}. Assuming $\R(t_{\xvec})$ is a well-behaved function, the approximation is still valid, but we have to modify Eq.~\eqref{eq:corr_2_point_1_bubble} to calculate the correlation of $\R$ directly instead of using $\delta t_{\xvec}$:
\begin{equation} \begin{split} \label{eq:2point_1_bubble_R}
&\int_{0}^{t_{\rm s}} E(\R_{\vec{x}} \R_{\vec{x}'}|\substack{\text{
bubble with } \\ \text{both points at t}})dP\left( \substack{\text{bubble with} \\ \text{both points at t}} \right) 	\\
&= \int_{0}^{t_{\rm s}} \, \R(t) ^2
\frac\Gamma V  V_{\text{overlap}}(t)dt.
\end{split}\end{equation} 
This result can be used to calculate the scalar spectrum for a general inflationary background, but only in the limit of small $\Gamma/V$.

Let us now consider a concrete example to demonstrate how the spectral shape  can be affected by the time dependence of $\frac{H}{\dot{\phi}^2}$.  In this example, $\frac{H}{\dot{\phi}^2}$ starts at some value $\frac{H}{\dot{\phi}_0^2}$, and remains constant until $t_{\rm drop}$, when it instantly changes to a new value smaller from the original one by a factor of $10$. In that scenario, the integrand of Eq.~\eqref{eq:R_integral} can be written as 
\begin{equation}
   \frac{H}{\dot{\phi}^2}=\frac{H}{\dot{\phi}_0^2}
   \left[\theta(t_{\rm drop}-t)+\frac{1}{10} \theta(t-t_{\rm drop})\right] ,
\end{equation}
and after integrating it, we get
\begin{equation} \begin{split}
	&\R(t_{\xvec}) = \\ &-\frac{H\Delta V_{\rm PT}}{\dot{\phi}_0^2}
\begin{cases}
	 t_{\xvec}-t_{\rm drop}+\frac{1}{10}(t_{\rm drop}-\mean{t_{\xvec}}), &t_{\xvec}< t_{\rm drop} \\
	 \frac{1}{10}(t_{\xvec}-\mean{t_{\xvec}}), &t_{\xvec}\ge t_{\rm drop} 
 \end{cases}.
\end{split}\end{equation}
The colored lines in Fig.~\ref{fig:P_R} were calculated by plugging this result into \eqref{eq:2point_1_bubble_R}.

\section{Non-Gaussianity and the induced gravitational waves} \label{app:GW}
Here, we very briefly review the equations necessary for calculating the secondary gravitational
waves, taken from \cite{Kohri_2018} and \cite{Adshead:2021hnm}. We then use the
methods of the previous section to calculate the four-point correlation
function and show that Eq.~\eqref{eq:P_h_integral} can be applied to our model,
ignoring non-Gaussianity.

\subsection{The induced gravitational waves}
To calculate the gravitational wave spectrum, we need the four-point correlation function of
$\R$. In the general, non-Gaussian case the correlation can be split into
disconnected and connected components \cite{Adshead:2021hnm}:
\begin{equation}
\begin{split}
	\mean{\mathcal{R}_{\vec{k}_{1}}
		\mathcal{R}_{\vec{k}_{2}} \mathcal{R}_{\vec{k}_{3}}
	\mathcal{R}_{\vec{k}_{4}}}=
&\left\langle\mathcal{R}_{\vec{k}_{1}}
	\mathcal{R}_{\vec{k}_{2}} \mathcal{R}_{\vec{k}_{3}}
\mathcal{R}_{\vec{k}_{4}}\right\rangle_{\mathrm{d}}+ \\
&\left\langle\mathcal{R}_{\vec{k}_{1}}
	\mathcal{R}_{\vec{k}_{2}} \mathcal{R}_{\vec{k}_{3}}
\mathcal{R}_{\vec{k}_{4}}\right\rangle_{\mathrm{c}}.
\end{split}
\end{equation}
The disconnected part satisfies Wick's theorem,
\begin{equation} \label{eq:diconnected_part}
\begin{split}
	\left\langle\mathcal{R}_{\vec{k}_{1}} \mathcal{R}_{\vec{k}_{2}}
	\mathcal{R}_{\vec{k}_{3}}
	\mathcal{R}_{\vec{k}_{4}}\right\rangle_{\mathrm{d}}=&\left\langle\mathcal{R}_{\vec{k}_{1}}
	\mathcal{R}_{\vec{k}_{2}}\right\rangle\left\langle\mathcal{R}_{\vec{k}_{3}}
	\mathcal{R}_{\vec{k}_{4}}\right\rangle+\\
	\left\langle\mathcal{R}_{\vec{k}_{2}}
	\mathcal{R}_{\vec{k}_{3}}\right\rangle\left\langle\mathcal{R}_{\vec{k}_{4}}
\mathcal{R}_{\vec{k}_{1}}\right\rangle 
&+\left\langle\mathcal{R}_{\vec{k}_{1}}
\mathcal{R}_{\vec{k}_{3}}\right\rangle\left\langle\mathcal{R}_{\vec{k}_{2}}
\mathcal{R}_{\vec{k}_{4}}\right\rangle\,,
\end{split}
\end{equation}
and the connected part defines the connected trispectrum $\mathcal{T}$,
\begin{equation}
	\begin{split}
	&\left\langle\mathcal{R}_{\vec{k}_{1}} \mathcal{R}_{\vec{k}_{2}} \mathcal{R}_{\vec{k}_{3}}
	\mathcal{R}_{\vec{k}_{4}}\right\rangle_{\mathrm{c}}= \\
	&\delta^{3}(\vec{k}_{1}+\vec{k}_{2}+\vec{k}_{3}+\vec{k}_{4})
	\mathcal{T}(\vec{k}_{1}, \vec{k}_{2}, \vec{k}_{3}, \vec{k}_{4})\,.
	\end{split}
\end{equation}
In the Gaussian case, the connected part vanishes because of Wick's theorem and
the four-point correlation function is fully described by the scalar spectrum
defined by Eq.~\eqref{eq:P_R_def}. 
	
The GW spectrum induced by the disconnected part is given by Eq.~\eqref{eq:P_h_integral},
where $\tilde{I}\left( v,u,k\eta \right) $ is a Green's function integral that
was calculated analytically in \cite{Kohri_2018}. 
Since we are only interested in the energy density of gravitational waves
today, we only need the late-time oscillation average of $\tilde{I}^2$, which, during radiation domination, is given by
\begin{equation} \label{eq:I_radiation}
	\begin{split}
		&\overline{\tilde{I}^2(v, u, k\eta \to \infty)}=
		\frac{1}{2}\left(\frac{3(u^{2}+v^{2}-3)}{4 u^{3} v^{3} k\eta}\right)^{2} \\
		&\Bigg[\left(-4 u v+\left(u^{2}+v^{2}-3\right) \log \left|\frac{3-(u+v)^{2}}{3-(u-v)^{2}}\right|\right)^{2} \\
		&+\pi^{2}\left(u^{2}+v^{2}-3\right)^{2} \Theta(v+u-\sqrt{3})\Bigg]\,.
	\end{split}
\end{equation}

The connected contribution to the GW spectrum is\footnote{Eq.~\eqref{eq:connected_GW} has a different
coefficient compared to equation (2.29) in \cite{Adshead:2021hnm} because here
we use the dimensionless spectrum.}
\begin{equation}
	\begin{split} \label{eq:connected_GW}
\left.\mathcal{P}_{\lambda}(k)\right|_{\mathrm{c}}=
	\frac{ k^3}{\pi^5}
	\int {\mathrm{d}^{3} q_{1}}
	\mathrm{d}^{3} q_{2} \,
	Q_{\lambda}\left(\vec{k}, \vec{q}_{1}\right) Q_{\lambda}\left(\vec{k},
	\vec{q}_{2}\right) \\
	I\left(\left|\vec{k}-\vec{q}_{1}\right|, q_{1},
\eta\right) I\left(\left|\vec{k}-\vec{q}_{2}\right|, q_{2}, \eta\right) \\
\times \mathcal{T}\left(\vec{q}_{1}, \vec{k}-\vec{q}_{1}, -\vec{q}_{2}, \vec{q}_2-\vec{k}\right)\,,
\end{split}
\end{equation}
where the $Q_{\lambda}$ are polarization factors, given by
\begin{equation}
Q_{\lambda}(\vec{k}, \vec{q}) \equiv \epsilon_{i j}^{\lambda}(\vec{k}) q_{i} q_{j},
\end{equation}
where $\epsilon_{ij}^{\lambda}$ with $\lambda = +,\times$ is a basis of traceless transverse polarization tensors. $I$ in Eq.~\eqref{eq:connected_GW} is related to $\tilde{I}$ used above through
a change of variables, $\tilde{I} \left( v,u,x \right) \equiv
k^2I\left(vk,uk,x/k\right) $. Taking $\vec{k}$ to be in the z direction and writing
$\vec{q}$ in spherical coordinates, $(q, \theta, \phi)$, the polarization factors can be written as
\begin{equation}
Q_{\lambda}(\vec{k}, \vec{q})=\frac{q^{2}}{\sqrt{2}} \sin ^{2}(\theta) \times\left\{\begin{array}{ll}
\cos (2 \phi) & \lambda=+ \\
\sin (2 \phi) & \lambda=\times.
\end{array}\right.
\end{equation}
Because of the polarization factors, the integral in Eq.~\eqref{eq:connected_GW}
does not vanish only when the connected trispectrum has a non-trivial
dependence on the azimuthal angles of $\vec{q}_1$ and $\vec{q}_2$.
In the next section, we will show the connected trispectrum in our model
does not depend on these angles. This is why the known result,
 Eq.~\eqref{eq:P_h_integral}, can be used on our model as if it
was Gaussian.

\subsection{The four-point correlation function}

We use the same method shown in App.~\ref{app:small_Gamma_2point} above for the two point correlation, and write an expression analogous to Eq.~\eqref{eq:corr_2_point_small_Gamma} for the 4-point correlation, assuming at most a single bubble (which is the leading contribution when expanding in small $\Gamma/V$),
\begin{equation} \begin{split} \label{eq:4_point_approx}
&\mean{\delta t_{\vec{x}_1} \delta t_{\vec{x}_2} \delta t_{\vec{x}_3} \delta t_{\vec{x}_4}}|_{\leq 1-{\rm bubble}} =  \\
	&E(\delta t_{\xvec_1}\delta t_{\xvec_2} \delta t_{\xvec_3} \delta t_{\xvec_4}|
 \substack{\text{bubble with} \\ \text{all pts before $t_{\rm s}$}})
 P(\substack{\text{bubble with} \\ \text{all pts before $t_{\rm s}$}}) \\
 +&\sum E(\delta t_{\xvec_1}\delta t_{\xvec_2} \delta t_{\xvec_3} \delta t_{\xvec_4}|
\substack{\text{bubble with only} \\ \text{3 pts before $t_{\rm s}$}})
 P(\substack{\text{bubble with only} \\ \text{3 pts before $t_{\rm s}$}}) \\
 +&\sum E(\delta t_{\xvec_1}\delta t_{\xvec_2} \delta t_{\xvec_3} \delta t_{\xvec_4}|
 \substack{\text{bubble with only} \\ \text{2 pts before $t_{\rm s}$}})
 P(\substack{\text{bubble with only} \\ \text{2 pts before $t_{\rm s}$}}) \\
 +&\sum E(\delta t_{\xvec_1}\delta t_{\xvec_2} \delta t_{\xvec_3} \delta t_{\xvec_4}|
 \substack{\text{bubble with only} \\ \text{1 pt before $t_{\rm s}$}})
 P(\substack{\text{bubble with only} \\ \text{1 pt before $t_{\rm s}$}}) \\
 +& E(\delta t_{\xvec_1}\delta t_{\xvec_2} \delta t_{\xvec_3} \delta t_{\xvec_4}|
 \substack{\text{no bubble} \\ \text{before $t_{\rm s}$}})
 P(\substack{\text{no bubble} \\ \text{before $t_{\rm s}$}})\,,
\end{split} \end{equation}
where $t_{\rm s}$ is now defined to be the moment after which no pair of two points is contained in a common Hubble sphere. The sums are over all possible choices of different points
to be included in the bubble.
As before, in the $\Gamma/V\to 0$ limit the first term is linear in $\Gamma/V$,
while the others are of order $\left(\Gamma/V\right)^2$ and above. This is because
the probability of forming a single bubble is linear in $\Gamma/V$, and
the expectation values that multiply them have a factor of
$\Gamma/V$ for every $\delta t_{\vec{x}}$  that doesn't tunnel before $t_{s}$.

The first term is given by an expression very similar to Eq.~\eqref{eq:corr_2_point_1_bubble},
\begin{equation}
\int_{0}^{t_{\rm s}} \, \left( t-\mean{t_{\xvec}} \right) ^4
\frac{\Gamma}{V} V_{\text{overlap}}(t)dt,
\end{equation} 
but this time
$V_{\text{overlap}}$ is the overlap volume between four Hubble spheres:
\begin{equation}
	\begin{split}
	V_{\text{overlap}}(t)=a^3(t)\int_{r_\mathcal{H}}^{} d^3y_1 \int_{r_\mathcal{H}}^{} d^3y_2 
	\int_{r_\mathcal{H}}^{} d^3y_3 \int_{r_\mathcal{H}}^{} d^3y_4 \\
	\delta\left( \vec{y}_1-\vec{y}_2+\vec{r}_2 \right) 
	\delta\left( \vec{y}_1-\vec{y}_3+\vec{r}_3 \right) 
	\delta\left( \vec{y}_1-\vec{y}_4+\vec{r}_4 \right) 
	\end{split}
\end{equation}
where $\vec{r}_i=\xvec_i-\xvec_1$ are the three independent separations between the $\xvec$ 's.
In a very similar manner to the above, we Fourier transform the volume to get
the contribution to the trispectrum, analogous to
Eq.~\eqref{eq:2_point_approx_final}. Omitting numerical coefficients, the result is
\begin{equation}
\begin{split}
	\mathcal{T}(\vec{k}_{1}, \vec{k}_{2}, \vec{k}_{3}, \vec{k}_{4}) \propto 
	\int_{0}^{\infty} dt \,\left( t-\mean{t_{\xvec}} \right) ^4 a^3(t) 
	\int_{r_\mathcal{H}}^{} d^3y_1e^{i\vec{k_1} \cdot\vec{y_1}} \\
	\int_{r_\mathcal{H}}^{} d^3y_2e^{i\vec{k_2} \cdot\vec{y_2}} 
	\int_{r_\mathcal{H}}^{} d^3y_3e^{i\vec{k_3} \cdot\vec{y_3}} 
	\int_{r_\mathcal{H}}^{} d^3y_4e^{i\vec{k_4} \cdot\vec{y_4}}. 
\end{split}
\end{equation} 
This expression depends only on the magnitudes of the $\vec{k}$ 's, not their
directions: $\mathcal{T}(\vec{k}_{1}, \vec{k}_{2}, \vec{k}_{3}, \vec{k}_{4})=
\mathcal{T}(k_{1}, k_{2}, k_{3}, k_{4})$. As mentioned above, this trispectrum
gives zero when plugged into Eq.~\eqref{eq:connected_GW}, because of the
integral over the azimuthal angle in the polarization factors.
This result has a physical interpretation: the leading contribution
we have calculated corresponds to the inhomogeneity created by the
presence of a single spherical bubble. A spherically symmetric inhomogeneity cannot emit gravitational waves. Contributions to the tunneling due to the formation of non-spherical bubbles may change this conclusion and consequently strengthen the predicted signal.  The study of such effects goes beyond the scope of this paper and is left for future work.

Since the leading term of order $\Gamma/V$ in Eq.~\eqref{eq:4_point_approx} does not
contribute, we have to calculate the $(\Gamma/V)^2$ terms.
In Eq.~\eqref{eq:4_point_approx}, only the second term is of that order.
However, taking the $(\Gamma/V)^{2}$ order means we have to
add terms with the probabilities that two bubbles
formed before $t_{\rm s}$, which were not present in Eq.~\eqref{eq:4_point_approx}. Since we are only interested in terms of order $(\Gamma/V)^2$, the
two bubbles have to cover all four points. We split this scenario into three cases:
\begin{enumerate}
	\item One of the bubbles includes all 4 points.
	\item One bubble includes a single point, and the other bubble includes the remaining the three points.
	\item Each bubble contains two points.
\end{enumerate}
We neglect the chance of two bubbles forming with a distance smaller than the Hubble radius, $H^{-1}$,
which means the three cases above are distinct.  This is justified since the mean distance between bubbles is of order $(\Gamma/V)^{-1/4} \gg H^{-1}$ for an incomplete PT.    

The first case has the same symmetry as in the single-bubble calculation above, and therefore 
gravitational waves are not produced. In the second case, after adding the second term
from Eq.~\eqref{eq:4_point_approx}, one of the points is independent of the 
other three, and since $\mean{\delta t_{\xvec}}=0$, the contribution to the
correlation function is zero. We are therefore only left with the last case,
where each of the two bubbles contains two points. The corresponding contribution 
to the four-point correlation function is then given by a sum over
the three possible ways of distributing the four points into two bubbles,
\begin{equation} \begin{split}
&\mean{\delta t_{\vec{x}_1} \delta t_{\vec{x}_2} \delta t_{\vec{x}_3} \delta t_{\vec{x}_4}} = \\
&\left(\frac\Gamma V\right)^2 \int_{0}^{t_{\rm s}} \left( t-\mean{t_{\xvec}} \right) ^2 V_{1,2}(t) dt
\int_{0}^{t_{\rm s}} \left( t-\mean{t_{\xvec}} \right) ^2 V_{3,4}(t) dt \\
+&\left(\frac\Gamma V\right)^2 \int_{0}^{t_{\rm s}} \left( t-\mean{t_{\xvec}} \right) ^2 V_{2,3}(t) dt
\int_{0}^{t_{\rm s}} \left( t-\mean{t_{\xvec}} \right) ^2 V_{1,4}(t) dt \\
+&\left(\frac\Gamma V\right)^2  \int_{0}^{t_{\rm s}} \left( t-\mean{t_{\xvec}} \right) ^2 V_{1,3}(t) dt
\int_{0}^{t_{\rm s}} \left( t-\mean{t_{\xvec}} \right) ^2 V_{2,4}(t) dt
\end{split}\end{equation} 
where $V_{i,j}$ is the overlap physical volume of two Hubble spheres centered
around $\xvec_{i}$ and $\xvec_j$. Each term is in fact a product of two 2-point
correlation functions as given by Eq.~\eqref{eq:corr_2_point_1_bubble}, so the
above correlation satisfies Eq.~\eqref{eq:diconnected_part}.
This is the promised result: the leading non-vanishing contribution
has only a disconnected component, so we can use Eq.~\eqref{eq:P_h_integral} safely.
%TC:endignore
\bibliography{HM}

\begin{thebibliography}{55}
\expandafter\ifx\csname natexlab\endcsname\relax\def\natexlab#1{#1}\fi
\expandafter\ifx\csname bibnamefont\endcsname\relax
  \def\bibnamefont#1{#1}\fi
\expandafter\ifx\csname bibfnamefont\endcsname\relax
  \def\bibfnamefont#1{#1}\fi
\expandafter\ifx\csname citenamefont\endcsname\relax
  \def\citenamefont#1{#1}\fi
\expandafter\ifx\csname url\endcsname\relax
  \def\url#1{\texttt{#1}}\fi
\expandafter\ifx\csname urlprefix\endcsname\relax\def\urlprefix{URL }\fi
\providecommand{\bibinfo}[2]{#2}
\providecommand{\eprint}[2][]{\url{#2}}

\bibitem[{\citenamefont{Abbott et~al.}(2016)}]{LIGOScientific:2016aoc}
\bibinfo{author}{\bibfnamefont{B.~P.} \bibnamefont{Abbott}}
  \bibnamefont{et~al.} (\bibinfo{collaboration}{LIGO Scientific, Virgo}),
  \bibinfo{journal}{Phys. Rev. Lett.} \textbf{\bibinfo{volume}{116}},
  \bibinfo{pages}{061102} (\bibinfo{year}{2016}), \eprint{1602.03837}.

\bibitem[{\citenamefont{Abbott et~al.}(2018)}]{KAGRA:2013rdx}
\bibinfo{author}{\bibfnamefont{B.~P.} \bibnamefont{Abbott}}
  \bibnamefont{et~al.} (\bibinfo{collaboration}{KAGRA, LIGO Scientific, Virgo,
  VIRGO}), \bibinfo{journal}{Living Rev. Rel.} \textbf{\bibinfo{volume}{21}},
  \bibinfo{pages}{3} (\bibinfo{year}{2018}), \eprint{1304.0670}.

\bibitem[{\citenamefont{Amaro-Seoane et~al.}(2017)\citenamefont{Amaro-Seoane,
  Audley, Babak, Baker, Barausse, Bender, Berti, Binetruy, Born, Bortoluzzi
  et~al.}}]{amaroseoane2017laser}
\bibinfo{author}{\bibfnamefont{P.}~\bibnamefont{Amaro-Seoane}},
  \bibinfo{author}{\bibfnamefont{H.}~\bibnamefont{Audley}},
  \bibinfo{author}{\bibfnamefont{S.}~\bibnamefont{Babak}},
  \bibinfo{author}{\bibfnamefont{J.}~\bibnamefont{Baker}},
  \bibinfo{author}{\bibfnamefont{E.}~\bibnamefont{Barausse}},
  \bibinfo{author}{\bibfnamefont{P.}~\bibnamefont{Bender}},
  \bibinfo{author}{\bibfnamefont{E.}~\bibnamefont{Berti}},
  \bibinfo{author}{\bibfnamefont{P.}~\bibnamefont{Binetruy}},
  \bibinfo{author}{\bibfnamefont{M.}~\bibnamefont{Born}},
  \bibinfo{author}{\bibfnamefont{D.}~\bibnamefont{Bortoluzzi}},
  \bibnamefont{et~al.}, \emph{\bibinfo{title}{Laser interferometer space
  antenna}} (\bibinfo{year}{2017}), \eprint{1702.00786}.

\bibitem[{\citenamefont{Luo et~al.}(2016)}]{TianQin:2015yph}
\bibinfo{author}{\bibfnamefont{J.}~\bibnamefont{Luo}} \bibnamefont{et~al.}
  (\bibinfo{collaboration}{TianQin}), \bibinfo{journal}{Class. Quant. Grav.}
  \textbf{\bibinfo{volume}{33}}, \bibinfo{pages}{035010}
  (\bibinfo{year}{2016}), \eprint{1512.02076}.

\bibitem[{\citenamefont{Crowder and Cornish}(2005)}]{Crowder:2005nr}
\bibinfo{author}{\bibfnamefont{J.}~\bibnamefont{Crowder}} \bibnamefont{and}
  \bibinfo{author}{\bibfnamefont{N.~J.} \bibnamefont{Cornish}},
  \bibinfo{journal}{Phys. Rev. D} \textbf{\bibinfo{volume}{72}},
  \bibinfo{pages}{083005} (\bibinfo{year}{2005}), \eprint{gr-qc/0506015}.

\bibitem[{\citenamefont{Harry et~al.}(2006)\citenamefont{Harry, Fritschel,
  Shaddock, Folkner, and Phinney}}]{Harry:2006fi}
\bibinfo{author}{\bibfnamefont{G.~M.} \bibnamefont{Harry}},
  \bibinfo{author}{\bibfnamefont{P.}~\bibnamefont{Fritschel}},
  \bibinfo{author}{\bibfnamefont{D.~A.} \bibnamefont{Shaddock}},
  \bibinfo{author}{\bibfnamefont{W.}~\bibnamefont{Folkner}}, \bibnamefont{and}
  \bibinfo{author}{\bibfnamefont{E.~S.} \bibnamefont{Phinney}},
  \bibinfo{journal}{Class. Quant. Grav.} \textbf{\bibinfo{volume}{23}},
  \bibinfo{pages}{4887} (\bibinfo{year}{2006}), \bibinfo{note}{[Erratum:
  Class.Quant.Grav. 23, 7361 (2006)]}.

\bibitem[{\citenamefont{Corbin and Cornish}(2006)}]{Corbin:2005ny}
\bibinfo{author}{\bibfnamefont{V.}~\bibnamefont{Corbin}} \bibnamefont{and}
  \bibinfo{author}{\bibfnamefont{N.~J.} \bibnamefont{Cornish}},
  \bibinfo{journal}{Class. Quant. Grav.} \textbf{\bibinfo{volume}{23}},
  \bibinfo{pages}{2435} (\bibinfo{year}{2006}), \eprint{gr-qc/0512039}.

\bibitem[{\citenamefont{Kramer and Champion}(2013)}]{Kramer:2013kea}
\bibinfo{author}{\bibfnamefont{M.}~\bibnamefont{Kramer}} \bibnamefont{and}
  \bibinfo{author}{\bibfnamefont{D.~J.} \bibnamefont{Champion}},
  \bibinfo{journal}{Class. Quant. Grav.} \textbf{\bibinfo{volume}{30}},
  \bibinfo{pages}{224009} (\bibinfo{year}{2013}).

\bibitem[{\citenamefont{Hobbs et~al.}(2010)\citenamefont{Hobbs, Archibald,
  Arzoumanian, Backer, Bailes, Bhat, Burgay, Burke-Spolaor, Champion, Cognard
  et~al.}}]{2010}
\bibinfo{author}{\bibfnamefont{G.}~\bibnamefont{Hobbs}},
  \bibinfo{author}{\bibfnamefont{A.}~\bibnamefont{Archibald}},
  \bibinfo{author}{\bibfnamefont{Z.}~\bibnamefont{Arzoumanian}},
  \bibinfo{author}{\bibfnamefont{D.}~\bibnamefont{Backer}},
  \bibinfo{author}{\bibfnamefont{M.}~\bibnamefont{Bailes}},
  \bibinfo{author}{\bibfnamefont{N.~D.~R.} \bibnamefont{Bhat}},
  \bibinfo{author}{\bibfnamefont{M.}~\bibnamefont{Burgay}},
  \bibinfo{author}{\bibfnamefont{S.}~\bibnamefont{Burke-Spolaor}},
  \bibinfo{author}{\bibfnamefont{D.}~\bibnamefont{Champion}},
  \bibinfo{author}{\bibfnamefont{I.}~\bibnamefont{Cognard}},
  \bibnamefont{et~al.}, \bibinfo{journal}{Classical and Quantum Gravity}
  \textbf{\bibinfo{volume}{27}}, \bibinfo{pages}{084013}
  (\bibinfo{year}{2010}).

\bibitem[{\citenamefont{Janssen et~al.}(2015)}]{Janssen:2014dka}
\bibinfo{author}{\bibfnamefont{G.}~\bibnamefont{Janssen}} \bibnamefont{et~al.},
  \bibinfo{journal}{PoS} \textbf{\bibinfo{volume}{AASKA14}},
  \bibinfo{pages}{037} (\bibinfo{year}{2015}), \eprint{1501.00127}.

\bibitem[{\citenamefont{Punturo et~al.}(2010)}]{Punturo:2010zz}
\bibinfo{author}{\bibfnamefont{M.}~\bibnamefont{Punturo}} \bibnamefont{et~al.},
  \bibinfo{journal}{Class. Quant. Grav.} \textbf{\bibinfo{volume}{27}},
  \bibinfo{pages}{194002} (\bibinfo{year}{2010}).

\bibitem[{\citenamefont{Reitze et~al.}(2019)}]{Reitze:2019iox}
\bibinfo{author}{\bibfnamefont{D.}~\bibnamefont{Reitze}} \bibnamefont{et~al.},
  \bibinfo{journal}{Bull. Am. Astron. Soc.} \textbf{\bibinfo{volume}{51}},
  \bibinfo{pages}{035} (\bibinfo{year}{2019}), \eprint{1907.04833}.

\bibitem[{\citenamefont{Arzoumanian et~al.}(2020)}]{NANOGrav:2020bcs}
\bibinfo{author}{\bibfnamefont{Z.}~\bibnamefont{Arzoumanian}}
  \bibnamefont{et~al.} (\bibinfo{collaboration}{NANOGrav}),
  \bibinfo{journal}{Astrophys. J. Lett.} \textbf{\bibinfo{volume}{905}},
  \bibinfo{pages}{L34} (\bibinfo{year}{2020}), \eprint{2009.04496}.

\bibitem[{\citenamefont{Guth}(1981)}]{Guth:1980zm}
\bibinfo{author}{\bibfnamefont{A.~H.} \bibnamefont{Guth}},
  \bibinfo{journal}{Phys. Rev. D} \textbf{\bibinfo{volume}{23}},
  \bibinfo{pages}{347} (\bibinfo{year}{1981}).

\bibitem[{\citenamefont{Linde}(1982)}]{Linde:1981mu}
\bibinfo{author}{\bibfnamefont{A.~D.} \bibnamefont{Linde}},
  \bibinfo{journal}{Phys. Lett. B} \textbf{\bibinfo{volume}{108}},
  \bibinfo{pages}{389} (\bibinfo{year}{1982}).

\bibitem[{\citenamefont{Akrami et~al.}(2020)}]{Planck:2018jri}
\bibinfo{author}{\bibfnamefont{Y.}~\bibnamefont{Akrami}} \bibnamefont{et~al.}
  (\bibinfo{collaboration}{Planck}), \bibinfo{journal}{Astron. Astrophys.}
  \textbf{\bibinfo{volume}{641}}, \bibinfo{pages}{A10} (\bibinfo{year}{2020}),
  \eprint{1807.06211}.

\bibitem[{\citenamefont{Ade et~al.}(2018)}]{BICEP2:2018kqh}
\bibinfo{author}{\bibfnamefont{P.~A.~R.} \bibnamefont{Ade}}
  \bibnamefont{et~al.} (\bibinfo{collaboration}{BICEP2, Keck Array}),
  \bibinfo{journal}{Phys. Rev. Lett.} \textbf{\bibinfo{volume}{121}},
  \bibinfo{pages}{221301} (\bibinfo{year}{2018}), \eprint{1810.05216}.

\bibitem[{\citenamefont{Komatsu et~al.}(2009)}]{WMAP:2008lyn}
\bibinfo{author}{\bibfnamefont{E.}~\bibnamefont{Komatsu}} \bibnamefont{et~al.}
  (\bibinfo{collaboration}{WMAP}), \bibinfo{journal}{Astrophys. J. Suppl.}
  \textbf{\bibinfo{volume}{180}}, \bibinfo{pages}{330} (\bibinfo{year}{2009}),
  \eprint{0803.0547}.

\bibitem[{\citenamefont{Carr et~al.}(2020)\citenamefont{Carr, Kohri, Sendouda,
  and Yokoyama}}]{Carr:2020gox}
\bibinfo{author}{\bibfnamefont{B.}~\bibnamefont{Carr}},
  \bibinfo{author}{\bibfnamefont{K.}~\bibnamefont{Kohri}},
  \bibinfo{author}{\bibfnamefont{Y.}~\bibnamefont{Sendouda}}, \bibnamefont{and}
  \bibinfo{author}{\bibfnamefont{J.}~\bibnamefont{Yokoyama}}
  (\bibinfo{year}{2020}), \eprint{2002.12778}.

\bibitem[{\citenamefont{{Fixsen} et~al.}(1996)\citenamefont{{Fixsen}, {Cheng},
  {Gales}, {Mather}, {Shafer}, and {Wright}}}]{1996ApJ...473..576F}
\bibinfo{author}{\bibfnamefont{D.~J.} \bibnamefont{{Fixsen}}},
  \bibinfo{author}{\bibfnamefont{E.~S.} \bibnamefont{{Cheng}}},
  \bibinfo{author}{\bibfnamefont{J.~M.} \bibnamefont{{Gales}}},
  \bibinfo{author}{\bibfnamefont{J.~C.} \bibnamefont{{Mather}}},
  \bibinfo{author}{\bibfnamefont{R.~A.} \bibnamefont{{Shafer}}},
  \bibnamefont{and} \bibinfo{author}{\bibfnamefont{E.~L.}
  \bibnamefont{{Wright}}}, \bibinfo{journal}{\apj}
  \textbf{\bibinfo{volume}{473}}, \bibinfo{pages}{576} (\bibinfo{year}{1996}),
  \eprint{astro-ph/9605054}.

\bibitem[{\citenamefont{Baumann}(2012)}]{baumann2012tasi}
\bibinfo{author}{\bibfnamefont{D.}~\bibnamefont{Baumann}},
  \emph{\bibinfo{title}{Tasi lectures on inflation}} (\bibinfo{year}{2012}),
  \eprint{0907.5424}.

\bibitem[{\citenamefont{An et~al.}(2020)\citenamefont{An, Lyu, Wang, and
  Zhou}}]{An:2020fff}
\bibinfo{author}{\bibfnamefont{H.}~\bibnamefont{An}},
  \bibinfo{author}{\bibfnamefont{K.-F.} \bibnamefont{Lyu}},
  \bibinfo{author}{\bibfnamefont{L.-T.} \bibnamefont{Wang}}, \bibnamefont{and}
  \bibinfo{author}{\bibfnamefont{S.}~\bibnamefont{Zhou}}
  (\bibinfo{year}{2020}), \eprint{2009.12381}.

\bibitem[{\citenamefont{Wang et~al.}(2019)\citenamefont{Wang, Cai, and
  Piao}}]{Wang:2018caj}
\bibinfo{author}{\bibfnamefont{Y.-T.} \bibnamefont{Wang}},
  \bibinfo{author}{\bibfnamefont{Y.}~\bibnamefont{Cai}}, \bibnamefont{and}
  \bibinfo{author}{\bibfnamefont{Y.-S.} \bibnamefont{Piao}},
  \bibinfo{journal}{Phys. Lett. B} \textbf{\bibinfo{volume}{789}},
  \bibinfo{pages}{191} (\bibinfo{year}{2019}), \eprint{1801.03639}.

\bibitem[{\citenamefont{An et~al.}(2022)\citenamefont{An, Lyu, Wang, and
  Zhou}}]{An:2022cce}
\bibinfo{author}{\bibfnamefont{H.}~\bibnamefont{An}},
  \bibinfo{author}{\bibfnamefont{K.-F.} \bibnamefont{Lyu}},
  \bibinfo{author}{\bibfnamefont{L.-T.} \bibnamefont{Wang}}, \bibnamefont{and}
  \bibinfo{author}{\bibfnamefont{S.}~\bibnamefont{Zhou}}
  (\bibinfo{year}{2022}), \eprint{2201.05171}.

\bibitem[{\citenamefont{Kosowsky and Turner}(1993)}]{Kosowsky:1992vn}
\bibinfo{author}{\bibfnamefont{A.}~\bibnamefont{Kosowsky}} \bibnamefont{and}
  \bibinfo{author}{\bibfnamefont{M.~S.} \bibnamefont{Turner}},
  \bibinfo{journal}{Phys. Rev. D} \textbf{\bibinfo{volume}{47}},
  \bibinfo{pages}{4372} (\bibinfo{year}{1993}), \eprint{astro-ph/9211004}.

\bibitem[{\citenamefont{Kosowsky et~al.}(1992)\citenamefont{Kosowsky, Turner,
  and Watkins}}]{Kosowsky:1991ua}
\bibinfo{author}{\bibfnamefont{A.}~\bibnamefont{Kosowsky}},
  \bibinfo{author}{\bibfnamefont{M.~S.} \bibnamefont{Turner}},
  \bibnamefont{and} \bibinfo{author}{\bibfnamefont{R.}~\bibnamefont{Watkins}},
  \bibinfo{journal}{Phys. Rev. D} \textbf{\bibinfo{volume}{45}},
  \bibinfo{pages}{4514} (\bibinfo{year}{1992}).

\bibitem[{\citenamefont{Kamionkowski et~al.}(1994)\citenamefont{Kamionkowski,
  Kosowsky, and Turner}}]{Kamionkowski:1993fg}
\bibinfo{author}{\bibfnamefont{M.}~\bibnamefont{Kamionkowski}},
  \bibinfo{author}{\bibfnamefont{A.}~\bibnamefont{Kosowsky}}, \bibnamefont{and}
  \bibinfo{author}{\bibfnamefont{M.~S.} \bibnamefont{Turner}},
  \bibinfo{journal}{Phys. Rev. D} \textbf{\bibinfo{volume}{49}},
  \bibinfo{pages}{2837} (\bibinfo{year}{1994}), \eprint{astro-ph/9310044}.

\bibitem[{\citenamefont{Huber and Konstandin}(2008)}]{Huber:2008hg}
\bibinfo{author}{\bibfnamefont{S.~J.} \bibnamefont{Huber}} \bibnamefont{and}
  \bibinfo{author}{\bibfnamefont{T.}~\bibnamefont{Konstandin}},
  \bibinfo{journal}{JCAP} \textbf{\bibinfo{volume}{09}}, \bibinfo{pages}{022}
  (\bibinfo{year}{2008}), \eprint{0806.1828}.

\bibitem[{\citenamefont{Caprini et~al.}(2008)\citenamefont{Caprini, Durrer, and
  Servant}}]{Caprini:2007xq}
\bibinfo{author}{\bibfnamefont{C.}~\bibnamefont{Caprini}},
  \bibinfo{author}{\bibfnamefont{R.}~\bibnamefont{Durrer}}, \bibnamefont{and}
  \bibinfo{author}{\bibfnamefont{G.}~\bibnamefont{Servant}},
  \bibinfo{journal}{Phys. Rev. D} \textbf{\bibinfo{volume}{77}},
  \bibinfo{pages}{124015} (\bibinfo{year}{2008}), \eprint{0711.2593}.

\bibitem[{\citenamefont{Caprini et~al.}(2016)}]{Caprini:2015zlo}
\bibinfo{author}{\bibfnamefont{C.}~\bibnamefont{Caprini}} \bibnamefont{et~al.},
  \bibinfo{journal}{JCAP} \textbf{\bibinfo{volume}{04}}, \bibinfo{pages}{001}
  (\bibinfo{year}{2016}), \eprint{1512.06239}.

\bibitem[{\citenamefont{Caprini et~al.}(2020)}]{Caprini:2019egz}
\bibinfo{author}{\bibfnamefont{C.}~\bibnamefont{Caprini}} \bibnamefont{et~al.},
  \bibinfo{journal}{JCAP} \textbf{\bibinfo{volume}{03}}, \bibinfo{pages}{024}
  (\bibinfo{year}{2020}), \eprint{1910.13125}.

\bibitem[{\citenamefont{Hindmarsh et~al.}(2014)\citenamefont{Hindmarsh, Huber,
  Rummukainen, and Weir}}]{Hindmarsh:2013xza}
\bibinfo{author}{\bibfnamefont{M.}~\bibnamefont{Hindmarsh}},
  \bibinfo{author}{\bibfnamefont{S.~J.} \bibnamefont{Huber}},
  \bibinfo{author}{\bibfnamefont{K.}~\bibnamefont{Rummukainen}},
  \bibnamefont{and} \bibinfo{author}{\bibfnamefont{D.~J.} \bibnamefont{Weir}},
  \bibinfo{journal}{Phys. Rev. Lett.} \textbf{\bibinfo{volume}{112}},
  \bibinfo{pages}{041301} (\bibinfo{year}{2014}), \eprint{1304.2433}.

\bibitem[{\citenamefont{Giblin and Mertens}(2014)}]{Giblin:2014qia}
\bibinfo{author}{\bibfnamefont{J.~T.} \bibnamefont{Giblin}} \bibnamefont{and}
  \bibinfo{author}{\bibfnamefont{J.~B.} \bibnamefont{Mertens}},
  \bibinfo{journal}{Phys. Rev. D} \textbf{\bibinfo{volume}{90}},
  \bibinfo{pages}{023532} (\bibinfo{year}{2014}), \eprint{1405.4005}.

\bibitem[{\citenamefont{Hindmarsh et~al.}(2015)\citenamefont{Hindmarsh, Huber,
  Rummukainen, and Weir}}]{Hindmarsh:2015qta}
\bibinfo{author}{\bibfnamefont{M.}~\bibnamefont{Hindmarsh}},
  \bibinfo{author}{\bibfnamefont{S.~J.} \bibnamefont{Huber}},
  \bibinfo{author}{\bibfnamefont{K.}~\bibnamefont{Rummukainen}},
  \bibnamefont{and} \bibinfo{author}{\bibfnamefont{D.~J.} \bibnamefont{Weir}},
  \bibinfo{journal}{Phys. Rev. D} \textbf{\bibinfo{volume}{92}},
  \bibinfo{pages}{123009} (\bibinfo{year}{2015}), \eprint{1504.03291}.

\bibitem[{\citenamefont{Kahniashvili et~al.}(2008)\citenamefont{Kahniashvili,
  Kosowsky, Gogoberidze, and Maravin}}]{Kahniashvili:2008pf}
\bibinfo{author}{\bibfnamefont{T.}~\bibnamefont{Kahniashvili}},
  \bibinfo{author}{\bibfnamefont{A.}~\bibnamefont{Kosowsky}},
  \bibinfo{author}{\bibfnamefont{G.}~\bibnamefont{Gogoberidze}},
  \bibnamefont{and} \bibinfo{author}{\bibfnamefont{Y.}~\bibnamefont{Maravin}},
  \bibinfo{journal}{Phys. Rev. D} \textbf{\bibinfo{volume}{78}},
  \bibinfo{pages}{043003} (\bibinfo{year}{2008}), \eprint{0806.0293}.

\bibitem[{\citenamefont{Kahniashvili et~al.}(2010)\citenamefont{Kahniashvili,
  Kisslinger, and Stevens}}]{Kahniashvili:2009mf}
\bibinfo{author}{\bibfnamefont{T.}~\bibnamefont{Kahniashvili}},
  \bibinfo{author}{\bibfnamefont{L.}~\bibnamefont{Kisslinger}},
  \bibnamefont{and} \bibinfo{author}{\bibfnamefont{T.}~\bibnamefont{Stevens}},
  \bibinfo{journal}{Phys. Rev. D} \textbf{\bibinfo{volume}{81}},
  \bibinfo{pages}{023004} (\bibinfo{year}{2010}), \eprint{0905.0643}.

\bibitem[{\citenamefont{Caprini et~al.}(2009)\citenamefont{Caprini, Durrer, and
  Servant}}]{Caprini:2009yp}
\bibinfo{author}{\bibfnamefont{C.}~\bibnamefont{Caprini}},
  \bibinfo{author}{\bibfnamefont{R.}~\bibnamefont{Durrer}}, \bibnamefont{and}
  \bibinfo{author}{\bibfnamefont{G.}~\bibnamefont{Servant}},
  \bibinfo{journal}{JCAP} \textbf{\bibinfo{volume}{12}}, \bibinfo{pages}{024}
  (\bibinfo{year}{2009}), \eprint{0909.0622}.

\bibitem[{\citenamefont{Schmitz}(2021)}]{Schmitz:2020syl}
\bibinfo{author}{\bibfnamefont{K.}~\bibnamefont{Schmitz}},
  \bibinfo{journal}{JHEP} \textbf{\bibinfo{volume}{01}}, \bibinfo{pages}{097}
  (\bibinfo{year}{2021}), \eprint{2002.04615}.

\bibitem[{\citenamefont{Matarrese et~al.}(1998)\citenamefont{Matarrese,
  Mollerach, and Bruni}}]{Matarrese:1997ay}
\bibinfo{author}{\bibfnamefont{S.}~\bibnamefont{Matarrese}},
  \bibinfo{author}{\bibfnamefont{S.}~\bibnamefont{Mollerach}},
  \bibnamefont{and} \bibinfo{author}{\bibfnamefont{M.}~\bibnamefont{Bruni}},
  \bibinfo{journal}{Phys. Rev. D} \textbf{\bibinfo{volume}{58}},
  \bibinfo{pages}{043504} (\bibinfo{year}{1998}), \eprint{astro-ph/9707278}.

\bibitem[{\citenamefont{Mollerach et~al.}(2004)\citenamefont{Mollerach, Harari,
  and Matarrese}}]{Mollerach:2003nq}
\bibinfo{author}{\bibfnamefont{S.}~\bibnamefont{Mollerach}},
  \bibinfo{author}{\bibfnamefont{D.}~\bibnamefont{Harari}}, \bibnamefont{and}
  \bibinfo{author}{\bibfnamefont{S.}~\bibnamefont{Matarrese}},
  \bibinfo{journal}{Phys. Rev. D} \textbf{\bibinfo{volume}{69}},
  \bibinfo{pages}{063002} (\bibinfo{year}{2004}), \eprint{astro-ph/0310711}.

\bibitem[{\citenamefont{Ananda et~al.}(2007)\citenamefont{Ananda, Clarkson, and
  Wands}}]{Ananda:2006af}
\bibinfo{author}{\bibfnamefont{K.~N.} \bibnamefont{Ananda}},
  \bibinfo{author}{\bibfnamefont{C.}~\bibnamefont{Clarkson}}, \bibnamefont{and}
  \bibinfo{author}{\bibfnamefont{D.}~\bibnamefont{Wands}},
  \bibinfo{journal}{Phys. Rev. D} \textbf{\bibinfo{volume}{75}},
  \bibinfo{pages}{123518} (\bibinfo{year}{2007}), \eprint{gr-qc/0612013}.

\bibitem[{\citenamefont{Baumann et~al.}(2007)\citenamefont{Baumann, Steinhardt,
  Takahashi, and Ichiki}}]{Baumann:2007zm}
\bibinfo{author}{\bibfnamefont{D.}~\bibnamefont{Baumann}},
  \bibinfo{author}{\bibfnamefont{P.~J.} \bibnamefont{Steinhardt}},
  \bibinfo{author}{\bibfnamefont{K.}~\bibnamefont{Takahashi}},
  \bibnamefont{and} \bibinfo{author}{\bibfnamefont{K.}~\bibnamefont{Ichiki}},
  \bibinfo{journal}{Phys. Rev. D} \textbf{\bibinfo{volume}{76}},
  \bibinfo{pages}{084019} (\bibinfo{year}{2007}), \eprint{hep-th/0703290}.

\bibitem[{\citenamefont{Kohri and Terada}(2018)}]{Kohri_2018}
\bibinfo{author}{\bibfnamefont{K.}~\bibnamefont{Kohri}} \bibnamefont{and}
  \bibinfo{author}{\bibfnamefont{T.}~\bibnamefont{Terada}},
  \bibinfo{journal}{Phys. Rev. D} \textbf{\bibinfo{volume}{97}},
  \bibinfo{pages}{123532} (\bibinfo{year}{2018}), \eprint{1804.08577}.

\bibitem[{\citenamefont{Coleman and De~Luccia}(1980)}]{ColemanDeLuccia}
\bibinfo{author}{\bibfnamefont{S.}~\bibnamefont{Coleman}} \bibnamefont{and}
  \bibinfo{author}{\bibfnamefont{F.}~\bibnamefont{De~Luccia}},
  \bibinfo{journal}{Phys. Rev. D} \textbf{\bibinfo{volume}{21}},
  \bibinfo{pages}{3305} (\bibinfo{year}{1980}).

\bibitem[{\citenamefont{Hawking and Moss}(1987)}]{Hawking:1981fz}
\bibinfo{author}{\bibfnamefont{S.~W.} \bibnamefont{Hawking}} \bibnamefont{and}
  \bibinfo{author}{\bibfnamefont{I.~G.} \bibnamefont{Moss}},
  \bibinfo{journal}{Adv. Ser. Astrophys. Cosmol.} \textbf{\bibinfo{volume}{3}},
  \bibinfo{pages}{154} (\bibinfo{year}{1987}).

\bibitem[{\citenamefont{Guth and Weinberg}(1983)}]{GUTH1983321}
\bibinfo{author}{\bibfnamefont{A.~H.} \bibnamefont{Guth}} \bibnamefont{and}
  \bibinfo{author}{\bibfnamefont{E.~J.} \bibnamefont{Weinberg}},
  \bibinfo{journal}{Nuclear Physics B} \textbf{\bibinfo{volume}{212}},
  \bibinfo{pages}{321} (\bibinfo{year}{1983}).

\bibitem[{\citenamefont{Turner et~al.}(1992)\citenamefont{Turner, Weinberg, and
  Widrow}}]{Turner:1992tz}
\bibinfo{author}{\bibfnamefont{M.~S.} \bibnamefont{Turner}},
  \bibinfo{author}{\bibfnamefont{E.~J.} \bibnamefont{Weinberg}},
  \bibnamefont{and} \bibinfo{author}{\bibfnamefont{L.~M.}
  \bibnamefont{Widrow}}, \bibinfo{journal}{Phys. Rev. D}
  \textbf{\bibinfo{volume}{46}}, \bibinfo{pages}{2384} (\bibinfo{year}{1992}).

\bibitem[{\citenamefont{Starobinsky}(1986)}]{Starobinsky:1986fx}
\bibinfo{author}{\bibfnamefont{A.~A.} \bibnamefont{Starobinsky}},
  \bibinfo{journal}{Lect. Notes Phys.} \textbf{\bibinfo{volume}{246}},
  \bibinfo{pages}{107} (\bibinfo{year}{1986}).

\bibitem[{\citenamefont{Linde}(1990)}]{Linde:2005ht}
\bibinfo{author}{\bibfnamefont{A.~D.} \bibnamefont{Linde}},
  \emph{\bibinfo{title}{{Particle physics and inflationary cosmology}}},
  vol.~\bibinfo{volume}{5} (\bibinfo{publisher}{CRC press},
  \bibinfo{year}{1990}), \eprint{hep-th/0503203}.

\bibitem[{\citenamefont{Chluba et~al.}(2012)\citenamefont{Chluba, Erickcek, and
  Ben-Dayan}}]{Chluba:2012we}
\bibinfo{author}{\bibfnamefont{J.}~\bibnamefont{Chluba}},
  \bibinfo{author}{\bibfnamefont{A.~L.} \bibnamefont{Erickcek}},
  \bibnamefont{and}
  \bibinfo{author}{\bibfnamefont{I.}~\bibnamefont{Ben-Dayan}},
  \bibinfo{journal}{Astrophys. J.} \textbf{\bibinfo{volume}{758}},
  \bibinfo{pages}{76} (\bibinfo{year}{2012}), \eprint{1203.2681}.

\bibitem[{\citenamefont{Dewdney et~al.}(2009)\citenamefont{Dewdney, Hall,
  Schilizzi, and Lazio}}]{5136190}
\bibinfo{author}{\bibfnamefont{P.~E.} \bibnamefont{Dewdney}},
  \bibinfo{author}{\bibfnamefont{P.~J.} \bibnamefont{Hall}},
  \bibinfo{author}{\bibfnamefont{R.~T.} \bibnamefont{Schilizzi}},
  \bibnamefont{and} \bibinfo{author}{\bibfnamefont{T.~J. L.~W.}
  \bibnamefont{Lazio}}, \bibinfo{journal}{Proceedings of the IEEE}
  \textbf{\bibinfo{volume}{97}}, \bibinfo{pages}{1482} (\bibinfo{year}{2009}).

\bibitem[{\citenamefont{Sathyaprakash and Schutz}(2009)}]{Sathyaprakash:2009xs}
\bibinfo{author}{\bibfnamefont{B.~S.} \bibnamefont{Sathyaprakash}}
  \bibnamefont{and} \bibinfo{author}{\bibfnamefont{B.~F.}
  \bibnamefont{Schutz}}, \bibinfo{journal}{Living Rev. Rel.}
  \textbf{\bibinfo{volume}{12}}, \bibinfo{pages}{2} (\bibinfo{year}{2009}),
  \eprint{0903.0338}.

\bibitem[{\citenamefont{Moore et~al.}(2015)\citenamefont{Moore, Cole, and
  Berry}}]{Moore:2014lga}
\bibinfo{author}{\bibfnamefont{C.~J.} \bibnamefont{Moore}},
  \bibinfo{author}{\bibfnamefont{R.~H.} \bibnamefont{Cole}}, \bibnamefont{and}
  \bibinfo{author}{\bibfnamefont{C.~P.~L.} \bibnamefont{Berry}},
  \bibinfo{journal}{Class. Quant. Grav.} \textbf{\bibinfo{volume}{32}},
  \bibinfo{pages}{015014} (\bibinfo{year}{2015}), \eprint{1408.0740}.

\bibitem[{\citenamefont{Ratzinger and Schwaller}(2021)}]{Ratzinger:2020koh}
\bibinfo{author}{\bibfnamefont{W.}~\bibnamefont{Ratzinger}} \bibnamefont{and}
  \bibinfo{author}{\bibfnamefont{P.}~\bibnamefont{Schwaller}},
  \bibinfo{journal}{SciPost Phys.} \textbf{\bibinfo{volume}{10}},
  \bibinfo{pages}{047} (\bibinfo{year}{2021}), \eprint{2009.11875}.

\bibitem[{\citenamefont{Adshead et~al.}(2021)\citenamefont{Adshead, Lozanov,
  and Weiner}}]{Adshead:2021hnm}
\bibinfo{author}{\bibfnamefont{P.}~\bibnamefont{Adshead}},
  \bibinfo{author}{\bibfnamefont{K.~D.} \bibnamefont{Lozanov}},
  \bibnamefont{and} \bibinfo{author}{\bibfnamefont{Z.~J.} \bibnamefont{Weiner}}
  (\bibinfo{year}{2021}), \eprint{2105.01659}.

\end{thebibliography}

%\begin{thebibliography}{99}
%
%
%
%\end{thebibliography}

\end{document}